\documentclass{scrartcl}
\usepackage[utf8]{inputenc}
\usepackage[T1]{fontenc}
\usepackage{graphicx}
\usepackage{subcaption}   
\usepackage{textcomp} 
\usepackage{amssymb,amsthm,amsmath}
\usepackage{color}
\usepackage[backend=bibtex,
    style=numeric,
    sorting=none,
    giveninits=true,   
    maxbibnames=99, 
    maxcitenames=1,
    minnames=1,  
    natbib           
]{biblatex}
\AtEveryBibitem{%
  \clearfield{issn}%
  \clearfield{month}%
  \clearfield{url}%
  \clearfield{isbn}%
}
\usepackage{upgreek}
\usepackage{makeidx}
\usepackage{hyperref}
\usepackage{nomencl} 
\usepackage{array} 
\usepackage{csquotes} 
\usepackage{siunitx} 

\graphicspath{figures/}

\title{Scale-Bridging Phase-Field Modeling of Microstructure Evolution by FE\textsuperscript{2} Computational Homogenization}
\author{Geralf Hütter, Fenil Lathiya, Vincent von Oertzen,\\ Bjoern Kiefer}
\date{\today}

\usepackage{nomencl}
\usepackage{accents}
\usepackage[normalem]{ulem}

\makenomenclature

\newcommand{\newsym}[4]{%
  \nomenclature[#4]{\ensuremath{#2}}{#3}%
  \newcommand{#1}{#2}%
}

\newcommand{\upmath}[1]{\mathrm{#1}}

\newcommand{\dutilde}[1]{\underaccent{\approx}{#1}}

\renewcommand*{\dot}[1]{%
  \accentset{\mbox{\large\bfseries .}}{#1}}

\newsym{\vnorm}{\vect{n}}{Normalenvektor}{n}
\newcommand{\norm}[1]{\left|#1\right|}
\newcommand{\surface}[1]{\partial{#1}}
\renewcommand{\d}{\upmath{d}}
\newcommand{\var}[2][]{\delta_{#1}{#2}}
\newcommand{\dV}{\;\d V}
\newcommand{\dsurf}{\;\d S}

\newcommand{\grad}[2][{}]{\vect{\nabla}_{\! #1}{#2}}

\renewcommand{\div}[2][{}]{\grad[#1]{}\sprod #2}

\newcommand{\Laplaceop}[2][{}]{\nabla^2_{#1}{#2}}

\newcommand{\domain}[1][{}]{B_{#1}}
\newcommand{\domainbound}[1][{}]{\partial\domain[#1]}

\newcommand{\tens}[1]{\undertilde{\boldsymbol #1}}

\newcommand{\Tens}[1]{\dutilde{\boldsymbol #1}}

\newcommand{\vect}[1]{\underline{\boldsymbol #1}}

\newcommand{\partderivf}[3][{}]{\frac{\partial^{#1}{#2}}{\partial{#3}^{#1}}}

\newcommand{\tprod}{\otimes}

\newcommand{\sprod}{\cdot}

\newcommand{\ssprod}{:}

\newcommand{\vbase}[1]{\vect{e}_{#1}}
\newcommand{\ex}{\vbase{x}}

\newsym{\tident}{\tens{I}}{Einheitstensor}{i}
\newsym{\Tidentfs}{\Tens{I}_{\upmath{FS}}}{vollst. symmetr. Einheitstensor 4. Stufe}{ifs}
\newsym{\Tidents}{\Tens{I}_{\upmath{S}}}{symmetrisierender Einheitstensor 4. Stufe}{ifs}
\newsym{\Tidenttransp}{\Tens{I}_{\upmath{T}}}{transponierender Einheitstensor 4. Stufe}{it}

\newcommand{\Lagrangemult}[1][{}]{\lambda^{#1}}
\newcommand{\vLagrangemult}[1][{}]{\vect{\lambda}^{#1}}

\newsym{\Lagrangefunc}{\mathcal{L}}{Lagrange-Funktion}{L}

\newcommand{\density}{\rho}

\newcommand{\rate}[1]{\dot{#1}}

\newsym{\stress}{\tens{\sigma}}{Innere Spannung}{s}

\newsym{\displ}{\vect{u}}{Verschiebungsvektor}{u}

\newsym{\strain}{\tens{\varepsilon}}{Dehnung}{e}

\newsym{\dissipation}{D}{Dissipation}{d}
\newsym{\dissipationpl}{\pi}{plast.\ Dissipation}{p}

\newsym{\innerenergy}{\Phi}{inner Energie}{F}

\newsym{\entropy}{\eta}{Entropy}{n}

\newsym{\entropyflux}{\vect{h}}{Entropiefluss}{h}
\newsym{\temperature}{T}{Temperature}{t}
\newsym{\Helmenergy}{\psi}{Helmholtz-Energie}{P}
\newsym{\Helmenergyref}{\Helmenergy_0}{Helmholtz-Energie}{P0}
\newsym{\ratepotential}{\Pi}{Ratenpotential}{P}
\newsym{\heatflux}{\vect{q}}{Wärmestrom}{qth}
\newsym{\thermalconduct}{\lambda}{Wärmeleitfähigkeit}{l}

\newsym{\entropyflow}{\vect{h}}{Entropiestrom}{h}
\newsym{\entropysource}{p_{\powerlocextvoltherm}}{Entropieproduction}{pr}
\newsym{\entropysourcespontan}{p_{\upmath{s}}}{spontane Entropieproduktion}{ps}
\newsym{\mechworkflux}{\vect{p}_{\upmath{m}}}{mechanische Arbeit}{pm}
\newsym{\intrvar}{h}{innere Variable}{h}
\newsym{\intrvarset}{\left\{h\right\}}{innere Variablen}{h}

\newsym{\intrvardriveforce}{Y}{Triebkräfte der inneren Variablen}{Y}

\newsym{\dissfunc}{\mathcal{D}}{Dissipationsfunktion}{D}
\newsym{\disspotential}{\Phi}{Dissipationspotential}{P}
\newcommand{\HamiltonLagrange}[1][{}]{\mathcal{L}_{#1}}

\newsym{\powerglobint}{\mathcal{P}_{\upmath{int}}}{Power of internal forces}{Pi}
\newsym{\powerglobext}{\mathcal{P}_{\upmath{ext}}}{Power of external forces}{Pe}
\newsym{\powerlocextvolmech}{{p}_{\upmath{m,int}}}{Power of external mechanical volume forces}{pm}
\newsym{\powerlocextvoltherm}{r}{Heat source per volume}{pth}
\newsym{\powerlocintmech}{{p}_{\upmath{m,int}}}{Power of internal mechanical volume forces}{pm}

\newsym{\orderparam}{\eta}{order parameter}{\eta}
\newsym{\phasestress}{\vect{\xi}}{micro stress vector}{n}
\newsym{\microforceint}{\pi}{internal microforce}{p}
\newsym{\microforceen}{\microforceint_\mathrm{en}}{energetic microforce}{pen}
\newsym{\microforcediss}{\microforceint_\mathrm{diss}}{energetic microforce}{pdiss}
\newsym{\microforceext}{\gamma}{external microforce}{g}
\newsym{\orderparammacro}{\bar{\orderparam}}{macroscopic order parameter}{\eta}
\newsym{\orderparamgradmacro}{\bar{\vect{K}}}{macroscopic gradient of order parameter}{K}
\newcommand{\orderparamgradmacrocomp}{\bar{K}}
\newsym{\phasestressmacro}{\bar{\phasestress}}{macroscopic stress vector}{p}

\newsym{\microforcemacro}{\bar{\microforceint}}{macroscopic internal microforce}{p}
\newcommand{\Lagrangemultorderparam}{\Lagrangemult[\orderparam]}
\newcommand{\Lagrangemultorderparamgrad}{\vLagrangemult[K]}
\newsym{\interfacetens}{\tens{\beta}}{interface scale tensor}{b}
\newsym{\interfaceparam}{\beta}{interface scale parameter}{b}
\newcommand{\energybarrier}{\Delta\psi}
\newcommand{\mobility}{M}
\newcommand{\orderparameq}{\orderparam_\mathrm{eq}}
\newcommand{\wallthickness}{\ell_\Gamma}
\newcommand{\intrevoltimescale}{\tau}
\newcommand{\straintr}[1][{}]{\strain^{\upmath{tr}}_{#1}}

\newsym{\yieldcond}{\Phi}{Fließbedingung}{f}
\newsym{\yieldstress}{\sigma_0}{yield stress}{s0}
\newsym{\plastmul}{\lambda}{plastischer Multiplikator}{l}
\newsym{\emod}{E^{\upmath{(m)}}}{Elastizitätsmodul}{e}
\newsym{\Poissrat}{\nu}{Querkontraktionszahl}{n}
\newsym{\compmod}{K}{Kompressionsmodul}{k}

\newsym{\tstiff}{\Tens{C}}{elastischer Steifigkeitstensor}{c} 
\newsym{\microcompmod}{\tilde{K}}{Mikrokompressionsmodul}{k}
\newsym{\tstiffhigherorder}{\tilde{\vect{C}}}{Mikrosteifigkeit höherer Ordnung}{c}
\newsym{\strainpl}{\strain_{\upmath{pl}}}{plastische Dehnung}{e}
\newsym{\strainratepl}{\dot{\strain}_{\upmath{pl}}}{plastische Dehnungrate}{e}
\newsym{\strainel}{\strain_{\upmath{el}}}{elastische Dehnung}{e}

\newcommand{\powerglobintmacro}{\bar{\mathcal{P}}_{\upmath{int}}}
\newcommand{\virtworkmacro}{\virtual{\bar{W}}}
\newcommand{\virtworkmacroext}{\virtworkmacro_{\upmath{ext}}}
\newcommand{\virtworkmacroint}{\virtworkmacro_{\upmath{int}}}
\newcommand{\virtworkspecificmacro}{\virtual{\bar{\mathcal{W}}}}

\newsym{\displapprox}{\tilde{\displ}}{lokale Verschiebungsapproximation}{u}
\newcommand{\fluctuation}[1]{\Delta{#1}}
\newsym{\velocityapprox}{\tilde{\velocity}}{lokale Geschwindigkeitssapproximation}{v}
\newcommand{\displmacrocomp}{U}
\newsym{\displmacro}{\vect{\displmacrocomp}}{makroskop.\ Verschiebung}{U}

\newsym{\velocitymacro}{\vect{V}}{makroskop.\ Geschwindigkeit}{V}

\newsym{\strainmacrocomp}{E}{makroskop.\ Dehnung}{E}
\newsym{\strainmacro}{\tens{\strainmacrocomp}}{makroskop.\ Dehnung}{E}

\newsym{\densitymacro}{\overline{\density}}{makroskop.\ Dehnung}{E}

\newcommand{\defratemicrocomp}{d}
\newsym{\defratemicro}{\tens{\defratemicrocomp}}{mikroskop.\ Deformationsgeschwindigkeit}{d}
\newsym{\displerror}{\Delta{\displ}}{lokale Verschiebungsdifferenz}{Du}
\newsym{\velocityerror}{\Delta{\velocity}}{lokale Geschwindigkeitsdifferenz}{Dv}

\newcommand{\locmacrocomp}{X}
\newsym{\locmacro}{\vect{\locmacrocomp}}{makroskop. Ort}{X}
\newcommand{\locmicrocomp}{x}
\newsym{\locmicro}{\vect{\locmicrocomp}}{mikroskop. Ort}{x}
\newcommand{\locmicrorelcomp}{\xi}
\newsym{\locmicrorel}{\vect{\locmicrorelcomp}}{mikroskop. Ort}{x}
\newcommand{\stressmacro}{\tens{\Sigma}}

\newcommand{\averop}[2][{}]{\left\langle#2\right\rangle_{#1}} 

\newcommand{\domainmacro}{\bar{\domain}}

\newcommand{\dVmacro}{\dV_{\locmacro}}

\newcommand{\domaincell}[1][{}]{\Delta V#1}

\newcommand{\domaincellbound}[1][{}]{\surface{\domaincell[]}#1}
\newcommand{\domaincellboundhalf}[1][{}]{\surface{\domaincell[]}^+#1}

\newcommand{\geommomcomp}{G}
\newsym{\geommom}{\tens{\geommomcomp}}{geometr. Moment}{G}
\newsym{\geommommom}{\Tens{G}^{\upmath{M}}}{geometr. Moment 4. Ordnung}{GM}

\newcommand{\gradmacro}[1]{\grad[\locmacro]{#1}}

\newcommand{\divmacro}[1]{\div[\locmacro]{#1}}

\newsym{\Helmenergymicro}{\psi}{lokale Helmholtz-Energie}{P}
\newsym{\Helmenergymacro}{\Psi}{makroskopische Helmholtz-Energie}{P}

\newcommand{\RVEedgelength}{\ell_{\mathrm{RVE}}}

\newcommand{\virtual}[1]{\delta#1}

\newcommand{\virtwork}{\virtual{W}}

\newcommand{\virtworkint}[1][{}]{\virtwork_{\upmath{int{#1}}}}

\newcommand{\elemind}{e}

\newcommand{\charelemsize}{h^{\elemind}}

\newcommand{\GPindex}{\alpha}

\DeclareMathOperator*{\Min}{Min}

\begin{document}

\maketitle

\begin{abstract}
Phase-field models have become a standard tool for simulating complex microstructure evolution in materials, but their application to engineering-scale components is often hindered by prohibitive computational costs arising from the need to resolve fine-scale features. To address this challenge, we propose a consistent homogenization framework for phase-field theory. By enforcing a Hill-Mandel-type condition of micro-homogeneity formulated in terms of Gurtin's microforces, a well-posed boundary value problem is derived for the representative volume element (RVE), establishing rigorous micro-macro relations for both the order parameter and its gradient.

The theory is implemented within a computational two-scale (FE\textsuperscript{2}) scheme and validated against direct numerical simulations. Two distinct examples are investigated: a minimal Allen-Cahn model and a mechanically-coupled model for stress-driven martensitic phase transformation. The results demonstrate that the proposed framework can reliably predict the spatial and temporal evolution of the macroscopically averaged fields with reasonable accuracy. 

\end{abstract}

\section{Introduction}

The phase-field method has emerged as a powerful and versatile tool in computational materials science with widespread usage. Its applications cover a vast range of phenomena, including phase transitions such as solidification \cite{schneider+etal2015, Heulens2011} and martensitic phase transformation \cite{Kochmann2016, levitas+preston2002, Rajendran2020, Stupkiewicz2002, von-oertzen+etal25}, the evolution of polycrystalline microstructures through grain boundary migration \cite{Ask2018}, and the modeling of material failure via damage and fracture mechanics \cite{Miehe2010, Kuhn2010, Azinpour2018}. More recently, its scope has expanded to complex multi-physics problems such as hydrogen-assisted material degradation \cite{martinez-paneda+golahmar+niordson18, svendsen+shanthraj+raabe2018, bartels+kurzeja+mosler2021, bai+etal2022, diddige+roth+kiefer25} and even to fields like topology optimization \cite{bourdin+chambolle06,Muench2018}. Comprehensive reviews by \citet{Chen2002a}, \citet{Moelans2008}, and \citet{Steinbach2009} summarize the extensive body of work in this area.

Conceptually, these models are typically classified based on the nature of the order parameter: the Allen-Cahn (or time-dependent Ginzburg-Landau) equation \cite{allen+cahn1979} is used for non-conserved quantities, while the Cahn-Hilliard equation \cite{cahn+hilliard1958} describes the evolution of conserved quantities. A key feature of these models is that the characteristic length scale, or the wavelength of fluctuations of the order parameter, is typically orders of magnitude smaller than the dimensions of an engineering structure. Consequently, a fully-resolved phase-field simulation of a complete component is computationally prohibitive in almost all cases.

To overcome this computational barrier, the principle of \emph{separation of scales} is commonly exploited through \emph{homogenization}, where the fields are split into a macroscopic average and a microscopic fluctuation. In computational mechanics, two-scale simulation techniques, such as the Finite Element Squared (FE\textsuperscript{2}) method or FE-FFT approaches, are now widely-established tools for handling this scale transition, as reviewed by \citet{Geers_multiscale_2017}. This field of research has recently seen further acceleration through machine learning-based approaches and model order reduction (MOR) techniques, with comparative studies highlighting their efficiency \cite{Rocha2020, Lange2025b}.

The success of this multiscale paradigm is not limited to purely mechanical problems but has been extended to a variety of coupled multi-field problems, including thermo-mechanical \cite{Oezdemir_thermo_mech_fe2_2008, berthelsen+menzel19}, electro-mechanical \cite{khalaquzzaman+etal12, keip+steinmann+schroeder14}, magneto-mechanical \cite{zabihyan+etal20}, and micromorphic problems \cite{Biswas2017, Rokos2018, Zhi2022, Malik2024}, and even to three-field settings \cite{schroeder+etal15, labusch+schroeder+keip18}.

Despite the maturity of two-scale methods and the prevalence of the phase-field method, a rigorous and consistent multiscale framework for the phase-field theory is still lacking. 
The term \enquote{homogenization} has been used in the phase-field context in a number of works \citep{Mosler2014,Schneider2015,Rancourt2016,kiefer+furlan+mosler17,Chatterjee2022}, among others, for the choice of constitutive interpolation schemes within diffuse interfaces to enforce mechanical compatibility, but without addressing multiscale applications.
A similar approach was used by \citet{Yu2023} for \enquote{regularization} within Gurtin's microforce concept, but none of these works constitute a multiscale framework. Our own recent attempts have concentrated on developing descriptors for phase morphology evolution, without the aim of establishing a rigorous kinetic scale transition \cite{Oertzen2024}.
\citet{Kochmann2016} used the phase-field theory at the RVE level, but included coupling to the macroscale via the mechanical fields only.

Multi-scale approaches have also been introduced in the context of the phase-field approach to fracture (PFA). Simpler models apply the PFA only at the macroscale, using a homogenized (undamaged) material behavior derived from the microscale \cite{Fantoni2019, He2020}. More advanced strategies employ established mechanical scale-bridging techniques but amend them with phase-field gradient effects in a more heuristic manner, without a rigorous coupling of the gradient terms \cite{Pise2024, Liu2025, Arunachala2024}. The most rigorous attempt to date \cite{Ma2024} utilizes asymptotic expansion of the linear phase-field PDE to derive an effective, tensorial macro-scale coefficient for the PFA's governing equation, thereby capturing potential anisotropy, but no nonlinearities.

\begin{figure}
    \centering
    \includegraphics{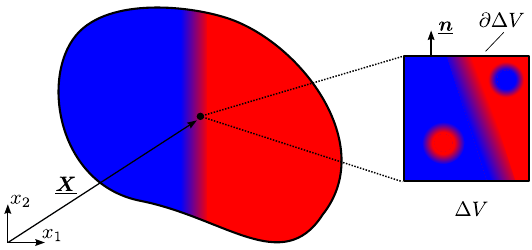}
    \caption{Visualization of a homogenization concept for the phase-field theory, in which an RVE $\domaincell$ is attached to each macroscopic point $\locmacro$ to resolve the fluctuations of the order parameter at the microscopic scale.}
    \label{fig:phase_field_homogenization}
\end{figure}

The present work aims to close this gap by proposing a consistent homogenization framework for phase-field theory as sketched in Figure~\ref{fig:phase_field_homogenization}, which overcomes the need to resolve all fluctuations of the order parameter at the macroscale, but uses effective macroscopic quantities instead. 
Our approach is developed within Gurtin's general framework of microforces and establishes a rigorous, thermodynamically-consistent transfer of both the phase-field and its gradient from the micro- to the macroscale, fully coupled with the mechanical fields. 
The resulting homogenization framework is implemented within a computational FE\textsuperscript{2} scheme to demonstrate its predictive capabilities, linking the macroscopic component behavior directly to the detailed evolution of the underlying microstructure.

\section{Phase-field theory}
\subsection{Microforce theory}

A pure dual phase-field theory introduces a continuous order parameter $\orderparam(\locmicro,t)$, characterizing the local phase state of the material, which is typically limited to a physically interpretable range, e.g.,$\orderparam\!\in\![-1,1]$. Its evolution is governed, in a variational description in terms of a stored energy density function $\Helmenergy$ and a dissipation potential $\disspotential$, neglecting external forces, by the condition \citep{tuma+etal2016}
\begin{equation}
    \ratepotential=\int\limits_{\domain}\rate{\Helmenergy}(\orderparam,\grad{\orderparam})+\disspotential(\rate{\orderparam})\dV\rightarrow \Min_{\rate{\orderparam}}
    \label{eq:variationalproblem}
\end{equation}
yielding Gurtin's microforce balance \citep{Gurtin1996}
\begin{equation}
   \div{\phasestress}+\microforceint=0\,.
   \label{eq:microforcebalance}
\end{equation}
Additionally, suitable boundary conditions have to be specified. In \eqref{eq:microforcebalance}, the microstress vector $\phasestress=\partial\Helmenergy/\partial\grad{\orderparam}$ and the microforce $\microforceint=\microforceen+\microforcediss$ appear, the latter comprising energetic and dissipative contributions, respectively,
\begin{align*}
   \microforceen&=-\,\partderivf{\Helmenergy}{\orderparam}\ , &
   \microforcediss&=-\,\partderivf{\disspotential}{\rate{\orderparam}}\,.
\end{align*}
In an Allen-Cahn framework, the potentials typically take the form $\Helmenergy=\Helmenergy_\mathrm{ch}(\orderparam)+\frac{\beta}{2}\left|\grad{\orderparam}\right|^2$ and $\disspotential=\frac{M}{2}\!\:\rate{\orderparam}^2$. 
The balance of energy then states that the internal power $\powerglobint$ equals the external power $\mathcal{P}_{\mathrm{ext}}$, which is zero in the absence of external forces and therefore $\powerglobint\!=\!0$. Here, the internal stress power is defined as
\begin{equation}
    \powerglobint=\int\limits_{\domain} \phasestress\sprod\grad{\rate{\orderparam}}  -\microforceint\rate{\orderparam} \dV
    \label{eq:intpower}
\end{equation}
(the minus in front of the last term is related to Gurtin's sign definition of $\microforceint$ in Eq.~\eqref{eq:microforcebalance}). Applying Gau\ss's divergence theorem leads to
\begin{equation}
    \int\limits_{\domain} \phasestress\sprod\grad{\rate{\orderparam}}  -\microforceint\rate{\orderparam} \dV=\int\limits_{\domainbound}\vnorm\sprod\phasestress\,\rate{\orderparam} \;\!\mathrm{d}A - \int\limits_{\domain}[\div{\phasestress} + \microforceint]\,\rate{\orderparam}\dV = 0 \ ,
\end{equation}
so that the balance of energy is directly fulfilled by \eqref{eq:variationalproblem} under the assumption of homogeneous Neumann boundary conditions.

\subsection{Hill's classical theory of mechanical homogenization}
The classical theory of mechanical homogenization, established by \citet{Hill1963}, forms the foundation for linking microscopic and macroscopic mechanical quantities in a consistent manner. Central to this theory are the \emph{micro-macro relations}, which define the macroscopic strain tensor $\strainmacro$ and stress tensor $\stressmacro$ as volumetric averages over a representative volume element $\domaincell$,
\begin{align}
    \strainmacro&=\averop{\strain}=\frac{1}{|\domaincell|}\int\limits_{\domaincell}\strain(\locmicro)\,\dV,\\
    \stressmacro&=\averop{\stress}=\frac{1}{|\domaincell|}\int\limits_{\domaincell}\stress(\locmicro)\,\dV.
\end{align}
Here and in the following, lower-case symbols denote microscopic quantities, while capital symbols, or symbols with an overbar, refer to the macroscale.
A cornerstone of Hill’s framework is the \emph{condition of macrohomogeneity} (Hill-Mandel principle), which ensures energetic equivalence between the microscopic and macroscopic descriptions. This condition requires that the volume average of the microscopic internal power equals the macroscopic internal power,
\begin{equation}
    \averop{\stress\ssprod\rate{\strain}}=\stressmacro\ssprod\rate{\strainmacro} \,.
    \label{eq:Hill-Mandel}
\end{equation}
This condition is satisfied for RVEs subjected to certain boundary conditions, such as
linear displacement (Dirichlet), uniform traction (Neumann), and periodic boundary conditions. 

\subsection{Homogenization for the phase-field theory}

Adopting the first part of Hill's concepts, the macroscopic kinematic  quantities  of the phase-field theory are defined as averages of their microscopic counterparts over the RVE
\begin{align}
    \orderparammacro&=\averop{\orderparam}, &
    \orderparamgradmacro&=\averop{\grad{\orderparam}}\,.
    \label{eq:micro-macro}
\end{align}
It can be shown that differentiating and averaging commute inside the domain \citep{Whitaker1967,mei+auriaul+ng1996, Forest1998,Forest2011}, so that $\orderparamgradmacro=\gradmacro{\orderparammacro}$. Note the notational difference between $\grad{\orderparam}$ defining the microscopic gradient of $\orderparam(\locmicro,t)$ w.r.t.~$\locmicro$ and $\gradmacro{\orderparammacro}$ its macroscopic counterpart, i.e., the gradient of $\orderparammacro(\locmacro,t)$ w.r.t.~$\locmacro$.
Based on the internal stress power expression \eqref{eq:intpower}, and in analogy to \eqref{eq:Hill-Mandel} in the mechanical case, the condition
\begin{align}
      \phasestressmacro\sprod\rate{\orderparamgradmacro}-\microforcemacro\rate{\orderparammacro}=\averop{\phasestress\sprod\grad{\rate{\orderparam}}-\microforceint\rate{\orderparam}}
    \label{eq:Hill-Mandel-phasefield}
\end{align}
 postulates the equivalence of averaged microscopic and macroscopic power, with macroscopic stresses $\phasestressmacro$ and $\microforcemacro$ to be defined. 
Provided this condition can be satisfied, the macroscopic stress power  takes the form
\begin{align}
  \powerglobintmacro=\int\limits_{\domainmacro} \phasestressmacro\sprod\gradmacro{\rate{\orderparammacro}}  -\microforcemacro\rate{\orderparammacro} \dVmacro\,.
  \label{eq:powermacro}
\end{align}
Requiring that $\powerglobintmacro$ is equal to some suitable external part for any values of $\rate{\orderparammacro}(\locmacro)$, in line with the Coleman-Noll argument or the principle of virtual power, implies the macroscopic balance
\begin{equation}
    \divmacro{\phasestressmacro}+\microforcemacro=0\,.
   \label{eq:microforcebalancemacro}
\end{equation}
A complete boundary value problem (BVP) at the macroscale additionally requires suitable boundary conditions, typically either by prescribing $\orderparammacro$ or $\phasestressmacro\sprod\vnorm$. 
Now the question on the formulation of the boundary value problem at the microscale needs to be addressed. 

\subsubsection*{Uniform microtraction boundary conditions}
As proposed by \citet{Miehe2007} for the conventional Cauchy theory of continuum mechanics, and also adopted to micromorphic theories \citep{Jaenicke2012,Huetter2017}, the BVP can be constructed by applying the kinematic micro-macro relations as global constraints to the underlying variational problem via Lagrange multipliers.
The idea of introducing such global multipliers in phase-field models has also appeared in other contexts, e.g., in the nonlocal Allen–Cahn theory of \citet{Rubinstein1992}, but without a homogenization setting. 
For the phase-field problem \eqref{eq:variationalproblem}, with micro-macro relations~\eqref{eq:micro-macro}, the Lagrangian reads:
\begin{equation}
    \HamiltonLagrange=\frac{1}{|\domaincell|}\int\limits_{\domaincell}\rate{\Helmenergy}(\orderparam,\grad{\orderparam})+\disspotential(\rate{\orderparam})\dV
    +\Lagrangemultorderparam\left[\averop{\rate{\orderparam}}-\rate{\orderparammacro}\right]
    -\Lagrangemultorderparamgrad\cdot\left[\averop{\grad{\rate{\orderparam}}}-\rate{\orderparamgradmacro}\right]
    \rightarrow \Min_{\rate{\orderparam},\Lagrangemultorderparam,\Lagrangemultorderparamgrad}
    \label{eq:variationalproblemminconstraint}
\end{equation}
The stationary condition w.r.t.~$\rate{\orderparam}$
\begin{align}
    \var{\HamiltonLagrange}=\averop{\phasestress\sprod\var{\grad{\rate{\orderparam}}}-\microforceint\var{\rate{\orderparam}}+\Lagrangemultorderparam\var{\rate{\orderparam}}-\Lagrangemultorderparamgrad\sprod\var{\grad{\rate{\orderparam}}}}=0
    \label{eq:Lagrangeminimumconstraint}
\end{align}
leads to the governing Euler-Lagrange equations 
\begin{align}
    \div{\phasestress}+\microforceint&=\Lagrangemultorderparam & \text{in } \domaincell \label{eq:microforcebalancemicro1}\\[.5ex]
    \phasestress\sprod\vnorm&=\Lagrangemultorderparamgrad \sprod\vnorm & \text{on } \domaincellbound \label{eq:naturalBCmicro}
\end{align}
at the microscale. Using the actual rate $\var{\rate{\orderparam}}=\rate{\orderparam}$ as  the test function in Eq.~\eqref{eq:Lagrangeminimumconstraint}
\begin{align}
    \averop{\phasestress\sprod\grad{\rate{\orderparam}}-\microforceint\,\var{\rate{\orderparam}}}= 
    \Lagrangemultorderparamgrad\sprod\averop{\grad{\rate{\orderparam}}}
    -\Lagrangemultorderparam\averop{\rate{\orderparam}}
\end{align}
allows, in view the condition of macrohomogeneity \eqref{eq:Hill-Mandel-phasefield}, to identify the Lagrange multipliers with the phase stress $\Lagrangemultorderparamgrad\!=\!\phasestressmacro$ and microforce $\Lagrangemultorderparam\!=\!\microforcemacro$, respectively.
Equation \eqref{eq:naturalBCmicro} is the pendant to the uniform traction boundary conditions in mechanics and might be called \emph{uniform microtraction boundary condition}. In addition, \eqref{eq:microforcebalancemicro1} shows that the macroscopic microforce $\Lagrangemultorderparam=\microforcemacro$ appears as an additional residual in the microscopic balance
\begin{align}
    \div{\phasestress}+\microforceint&=\microforcemacro& \text{in } \domaincell \label{eq:microforcebalancemicro}
\end{align}
to ensure macroscopic compatibility of the order parameter on average.
Notably, $\microforcemacro$ vanishes, according to the macroscopic balance \eqref{eq:microforcebalancemacro}, if the coupling to the macroscopic gradient of the order parameter $\orderparamgradmacro$ in \eqref{eq:micro-macro}$_2$ is absent, implying $\phasestressmacro\!=\!0$. This decoupling corresponds to the approach of \citet{Kochmann2016}, where the microscopic order parameter field was computed using the original Allen-Cahn equation~\eqref{eq:microforcebalance} and the macroscopic order parameter field $\orderparammacro$ could have been obtained via post-processing. In this sense, the approach of \citet{Kochmann2016} can be termed a zeroth-order phase-field homogenization theory, whereas the present theory, which retains the macroscopic gradient, constitutes a first-order phase-field homogenization theory.

\subsubsection*{Linear order parameter boundary conditions}
Alternatively, \emph{linear order parameter boundary conditions}
\begin{align}
    \orderparam&=\orderparammacro+\orderparamgradmacro\sprod\left(\locmicro-\locmacro\right) & \text{on } \domaincellbound \label{eq:essentialBCmicro}
\end{align}
can be formulated, satisfying the kinematic relation \eqref{eq:micro-macro}$_2$ for the macroscopic gradient $\orderparamgradmacro$ immediately, if the macroscopic location $\locmacro$ is identified with the geometric center $\averop{\locmicro}$ of the RVE. 
The term $\orderparammacro$ is included on the right-hand side of \eqref{eq:essentialBCmicro} firstly to have a homogeneous field $ \orderparam(\locmicro)=\orderparammacro$ as a kinematically admissible solution and secondly for \eqref{eq:essentialBCmicro} to be the solution of Taylor-Voigt type, if prescribed in the whole domain $\domaincell$.
In general, however, \eqref{eq:essentialBCmicro} does not satisfy constraint \eqref{eq:micro-macro}$_1$ for the order parameter $\orderparammacro$ ad-hoc. Consequently, \eqref{eq:micro-macro}$_1$ still has to be ensured by a Lagrange multiplier, i.e.
\begin{equation}
    \HamiltonLagrange=\frac{1}{|\domaincell|}\int\limits_{\domaincell}\rate{\Helmenergy}(\orderparam,\grad{\orderparam})+\disspotential(\rate{\orderparam})\dV
    +\Lagrangemultorderparam\left[\averop{\rate{\orderparam}}-\rate{\orderparammacro}\right]
    \rightarrow \Min_{\rate{\orderparam},\Lagrangemultorderparam}
    \label{eq:variationalproblemlinorderparamBC}
\end{equation}
so that \eqref{eq:microforcebalancemicro} remains valid. The right-hand side of the condition of macrohomogeneity \eqref{eq:Hill-Mandel-phasefield} can be integrated by parts to yield
\begin{align}
    \phasestressmacro\sprod\rate{\orderparamgradmacro}-\microforcemacro\,\rate{\orderparammacro}=
    \frac{1}{|\domaincell|}\int\limits_{\domaincellbound}\rate{\orderparam}\,\phasestress\sprod\vnorm \dsurf
    -\averop{\rate{\orderparam}\left[\div{\phasestress}+\microforceint\right]}\,.
    \label{eq:Hill-Mandel-phasefield__lin}
\end{align}
Given the micro-macro relations \eqref{eq:micro-macro} and inserting boundary condition \eqref{eq:essentialBCmicro} into the first term as well as the micro-force balance \eqref{eq:microforcebalancemicro} into the second term therefore results in 
\begin{align}
    \phasestressmacro\sprod\rate{\orderparamgradmacro}-\microforcemacro\,\rate{\orderparammacro}   = \left[\frac{1}{|\domaincell|}\int\limits_{\domaincellbound}\left(\locmicro-\locmacro\right)\tprod\phasestress\sprod\vnorm \dsurf\right]\sprod\rate{\orderparamgradmacro} - \averop{\microforceint}\rate{\orderparammacro} \: .
\end{align}
The condition of macrohomogeneity can thus be satisfied if the macroscopic phase stress and microforce are defined as 
\begin{align}
    \microforcemacro&=\averop{\microforceint}\ , & \phasestressmacro&=\frac{1}{|\domaincell|}\int\limits_{\domaincellbound}\left(\locmicro-\locmacro\right)\tprod\phasestress\sprod\vnorm     \dsurf \ .
    \label{eq:micro-macrokinetics}
\end{align}
It can easily be verified that these relations are also satisfied by the uniform microtraction BCs \eqref{eq:naturalBCmicro}, in conjunction with Eq.~\eqref{eq:microforcebalancemicro}.

\subsubsection*{Periodic boundary conditions}
The third common option are \emph{periodic boundary conditions}. They add a fluctuation term $\fluctuation{\orderparam}$ to the linear order parameter boundary conditions \eqref{eq:essentialBCmicro}, i.e.,
\begin{align}
    \orderparam&=\orderparammacro+\orderparamgradmacro\sprod\left(\locmicro-\locmacro\right)+\fluctuation{\orderparam} & \text{on } \domaincellbound \ ,\label{eq:periodicBC}
\end{align}
which is restricted to be identical $\fluctuation{\orderparam}(\locmicro^+)=\fluctuation{\orderparam}(\locmicro^-)$ at homologous points $\locmicro^+$ and $\locmicro^-$ of the boundary at which $\vnorm(\locmicro^-)=-\,\vnorm(\locmicro^+)$. These conditions can equivalently be written as
\begin{equation}
    \orderparam(\locmicro^+)-\orderparam(\locmicro^-)=\orderparamgradmacro\sprod\left[\locmicro^+-\locmicro^-\right]\,. 
    \label{eq:periodicBCimpl}
\end{equation}
The respective Lagrangian functional for the rate potential reads
\begin{equation}
\begin{split}    
    \HamiltonLagrange=&\frac{1}{|\domaincell|}\int\limits_{\domaincell}\rate{\Helmenergy}(\orderparam,\grad{\orderparam})+\disspotential(\rate{\orderparam})\dV
    +\Lagrangemultorderparam\left[\averop{\rate{\orderparam}}-\rate{\orderparammacro}\right]\\
    &-\int\limits_{\domaincellboundhalf}\Lagrangemult(\locmicro^+)\left[\rate{\orderparam}(\locmicro^+)-\rate{\orderparam}(\locmicro^-)-\rate{\orderparamgradmacro}\sprod\left(\locmicro^+-\locmicro^-\right)\right]\dsurf
    \rightarrow \Min_{\rate{\orderparam},\Lagrangemultorderparam,\Lagrangemult(\locmicro^+)} .
\end{split}
    \label{eq:variationalproblemperiodic}
\end{equation}
Therein, $\locmicro^-$ is to be considered as a function of $\locmicro^+$, over which the surface integral is to be computed. The respective Euler-Lagrange equations are the microforce balance \eqref{eq:microforcebalancemicro} in the domain and at the boundary
\begin{equation}
    \phasestress\cdot\vnorm=
    \begin{cases}
        \phantom{-}\Lagrangemult(\locmicro^+) & \text{for } \locmicro\in\domaincellboundhalf\\
        -\Lagrangemult(\locmicro^+) & \text{else}
    \end{cases}\ .
\end{equation}
These antiperiodic \emph{microtractions}, together with \eqref{eq:periodicBCimpl}, satisfy the condition of macrohomogeneity for the given micro-macro relations \eqref{eq:micro-macrokinetics} in terms of the phase stress and the microforce.

\subsection{Numerical Implementation}
\label{sec:numericalimplementation}

The numerical implementation of the proposed multi-scale phase-field formulation requires solving the microscopic balance~\eqref{eq:microforcebalancemicro} as well as $\div{\stress}=0$, using the respective boundary conditions at both scales, concurrently with their macroscopic counterparts \eqref{eq:microforcebalancemacro} and $\divmacro{\stressmacro}=0$. A classical approach to achieve this is the FE\textsuperscript{2} method, as reviewed comprehensively by~\citet{Schroeder2014} and \citet{Matous2017}. In recent years, alternative techniques such as FE-FFT have gained increasing attention in the context of computational homogenization, as highlighted in the review by~\citet{Gierden2022}.

In the present contribution, however, the focus lies primarily on the theoretical framework rather than the development of numerical methods. For reasons of simplicity and compatibility with established multi-scale strategies, the FE\textsuperscript{2} methodology is adopted here as the numerical backbone of the approach.

At the macroscale, the principle of virtual power is formulated as
\begin{equation}
    \virtworkmacro = \virtworkmacroint - \virtworkmacroext = 0\ ,
\end{equation}
where the internal contribution $\virtworkmacroint$ is evaluated numerically as
\begin{equation}
    \virtworkmacroint = \sum_{\alpha} w_\alpha J_{\alpha}\,\virtworkspecificmacro_{\alpha}\ .
    \label{eq:virtpowerFE}
\end{equation}
Here, the summation extends over all integration points $\alpha$ of the macroscopic finite elements, with weights $w_{\alpha}$ and Jacobian determinants $J_{\alpha}$. The internal virtual power density $\virtworkspecificmacro$ depends on the underlying physical model. In the classical mechanical case, it takes the form $\virtworkspecificmacro \!=\! \stressmacro : \delta\strainmacro$, while in the (pure) phase-field context it reads $\virtworkspecificmacro = \phasestressmacro \cdot \delta\orderparamgradmacro - \microforcemacro \,\delta\orderparammacro$ according to~\eqref{eq:powermacro}. For coupled simulations involving both mechanics and phase-field evolution, these expressions are naturally combined.

In a traditional FE\textsuperscript{2} framework, a separate finite element simulation is performed at each macroscopic integration point to solve the microscopic problem. This procedure yields the macroscopic stresses $\stressmacro$, $\phasestressmacro$, and $\microforcemacro$ corresponding to prescribed macroscopic kinematic quantities $\strainmacro$, $\orderparamgradmacro$, and $\orderparammacro$. Furthermore, the condensed stiffness matrix is computed and transferred back to the macroscale. Iterative solutions are then performed at both scales, in a nested fashion, until convergence is reached simultaneously.

An alternative implementation strategy, known the DirectFE\textsuperscript{2} method, was proposed by~\citet{Tan2020}. In this approach, the condition of macrohomogeneity, Eqs.~\eqref{eq:Hill-Mandel} or  \eqref{eq:Hill-Mandel-phasefield}, respectively, is incorporated in variational form directly into the macroscopic principle of virtual power~\eqref{eq:virtpowerFE}. Specifically, the internal virtual work is expressed as
\begin{equation}
    \virtworkmacroint = \sum_{\alpha} \frac{w_\alpha J_\alpha}{\domaincell_\alpha}\:\virtworkint[,\GPindex]\ ,
    \qquad \text{with} \quad
    \virtworkint[,\GPindex] = \int\limits_{\domaincell_\alpha} \tens{\sigma} : \delta\tens{\varepsilon}+\phasestress\sprod\grad{\var{\orderparam}}-\microforceint\,\var{\orderparam}\;\mathrm{d}V\ ,
    \label{eq:internalWorkinserted}
\end{equation}
where $\virtworkint[,\GPindex]$ denotes the contribution to the virtual work integral by the domain $\domaincell_\alpha$ of the representative volume element that is associated with the macroscopic integration point $\alpha$. In practice, this integral is computed numerically after an FE discretization of the RVE. Consequently, the FE\textsuperscript{2} problem can be reformulated such that the discretized RVE is directly embedded into the macroscopic finite element model. This requires the introduction of appropriate kinematic constraints to link the microscopic nodal degrees of freedom (DOFs) $(\displ, \orderparam)$ with their macroscopic counterparts $(\displmacro, \orderparammacro)$ via the respective microscale boundary conditions. 
Note that the sum over $\alpha$ in \eqref{eq:internalWorkinserted} is taken over all macroscopic Gauss points of the mesh.

A crucial condition for this embedding is that the pre-factor in~\eqref{eq:internalWorkinserted}$_1$ equals unity. 
This ensures that the RVE virtual work replaces the macroscopic stress power contribution at each integration point on a one-to-one basis, without additional scaling. 
In the original work of~\citet{Tan2020}, this condition is satisfied by scaling the unit cell volume $\domaincell_\alpha$ to match the equivalent volume $w_\alpha J_\alpha$ of the corresponding macroscopic integration point. In the present formulation, however, such scaling is not feasible due to inherent size effects at the microscale associated with the gradient of the order parameter $\grad{\orderparam}$. 
So instead, we follow the alternative strategy proposed by \citet{Zhi2022} and extended by \citet{Malik2024}. This method is specifically designed for, and restricted to, plane problems, as it achieves the necessary scaling by adjusting the (nominal) out-of-plane thickness $t_{\alpha}$ of the microscale elements, while the physically relevant in-plane dimensions are kept constant.

In summary, the DirectFE\textsuperscript{2} approach eliminates the explicit computation of macroscopic stresses $\stressmacro_\alpha$, $\microforcemacro_\alpha$, and $\phasestressmacro_\alpha$ from the governing discretized equations by direct insertion of the macroscopic local virtual power at each integration point. Consequently, the macroscopic finite elements effectively degenerate into kinematic constraints between the macroscopic nodal DOFs and the macroscopic Gauss point quantities $\strainmacro_\alpha$, $\orderparammacro_\alpha$, and $\orderparamgradmacro_\alpha$. Additional kinematic constraints of those quantities to the RVE are necessary according to the type of boundary conditions at the microscale. All these linear constraints can be implemented straightforwardly in a commercial finite element software. In the present work, this is achieved in Simulia Abaqus, using an extended version of the \texttt{Python} script package developed by~\citet{Malik2024}.

Finally, the microscopic problem~\eqref{eq:microforcebalancemicro} is solved by exploiting its formal analogy to a heat transfer problem \citep{Levitas_2010}, where $\microforceint$ plays the role of a volumetric heat source and $\phasestress$ corresponds to the heat flux. This analogy permits leveraging \texttt{Abaqus} built-in thermal or thermomechanical elements for an efficient solution of the microscale equations---as described in detail below for the particular model. A similar strategy was demonstrated in~\cite{Seupel2018,Azinpour2018} for gradient-enhanced damage models governed by structurally equivalent PDEs.

\section{Examples}
\subsection{Allen-Cahn theory with double-well potential}

\subsubsection{Overview}

As a first example, we consider a minimal Allen-Cahn model. The model is defined by a bi-quadratic, symmetric double-well potential for the Helmholtz free energy, an isotropic gradient contribution, and a quadratic dissipation potential:
\begin{align}
    \Helmenergy(\orderparam,\grad{\orderparam}) &= \energybarrier\left[1-\orderparam^2\right]^2 + \frac{\interfaceparam}{2}\norm{\grad{\orderparam}}^2 \ ,\label{eq:freeenergyAllenCahn}\\
    \disspotential(\rate{\orderparam})&=\frac{1}{2\mobility}\,\rate{\orderparam}^2 \label{eq:DissPotAllenCahn} \,.
\end{align}
This formulation represents a common choice for many fundamental studies of phenomena describable by phase-field theory \cite{Chen2002a, Steinbach2009, Rancourt2016}. In these equations, the symbol $\energybarrier$ represents the energy barrier between the two equivalent equilibrium states $\orderparameq\!=\!\pm1$, the parameter $\interfaceparam$ penalizes the gradient of the order parameter according to the diffuse interface approach, and $\mobility$ is the (inverse) mobility that governs the kinetics of the phase evolution.

Substituting these potentials into the microforce balance, Eq.~\eqref{eq:microforcebalance}, yields the classical Allen-Cahn evolution equation:
\begin{align}
  \rate{\orderparam}=-\,\mobility\left[\partderivf{\Helmenergy}{\orderparam}-\interfaceparam\,\Laplaceop{\orderparam}\right]\,.
  \label{eq:AllenCahnequation}
\end{align}
The presence of the gradient parameter $\interfaceparam$ introduces an intrinsic length scale, $\wallthickness\!:=\!\Delta\orderparameq\sqrt{\frac{\interfaceparam}{\energybarrier}}$, see~ \cite{AllenCahn1979}, which characterizes the thickness of the diffuse interface. Similarly, the mobility $\mobility$ defines an intrinsic time scale for the evolution, $\intrevoltimescale\!:=\!\frac{1}{\mobility\energybarrier}$. The following results are presented in a normalized form with respect to these intrinsic scales, $\wallthickness$ and $\intrevoltimescale$, in order to be independent of a particular choice of constitutive parameters.

As mentioned, the numerical implementation of the homogenization framework is carried out in the commercial finite element software Abaqus, utilizing a custom DirectFE\textsuperscript{2} scheme as outlined in Section~\ref{sec:numericalimplementation}, making use of the formal analogy between the phase-field equations and the equations of heat transfer. Specifically, the order parameter $\orderparam$ is mapped to temperature, the gradient parameter $\interfaceparam$ to heat conductivity, and the inverse mobility $\mobility^{-1}$ to specific heat capacity. 
The source term $\partderivf{\Helmenergy}{\orderparam}$ in the evolution equation corresponds to a volumetric heat source, which is implemented via Abaqus' user subroutine interface \texttt{HETVAL}.
At the microscale, the representative volume element is discretized using the software's built-in eight-node quadrilateral heat transfer elements (referred to as \texttt{DC2D8} in the Abaqus documentation). The same shape functions are used to discretize the macroscopic problem. A quadratic RVE with an edge length of $\RVEedgelength$ is employed, discretized with a regular grid of elements $\charelemsize$. Unless stated otherwise, the element size is chosen as $\charelemsize\!=\!\frac{1}{20}\RVEedgelength$, based on a convergence study, which is not shown here for brevity. As usual, the nodal values of $\orderparam$ are initialized by stochastic noise, where the same initialization field is used for all RVEs within a single FE\textsuperscript{2} simulation.

\subsubsection{Type of boundary conditions}
\begin{figure}[hbt] 
\centering
\begin{tabular}{ m{1cm} m{0.8\textwidth} }
 UBC & \includegraphics[width=\linewidth]{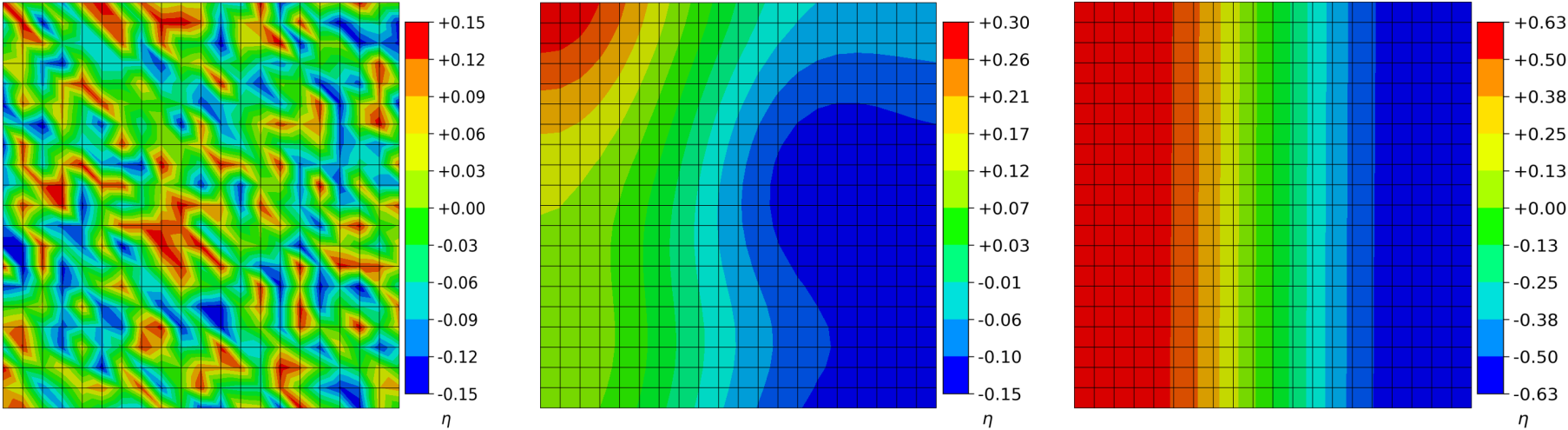}\\
 LBC & \includegraphics[width=\linewidth]{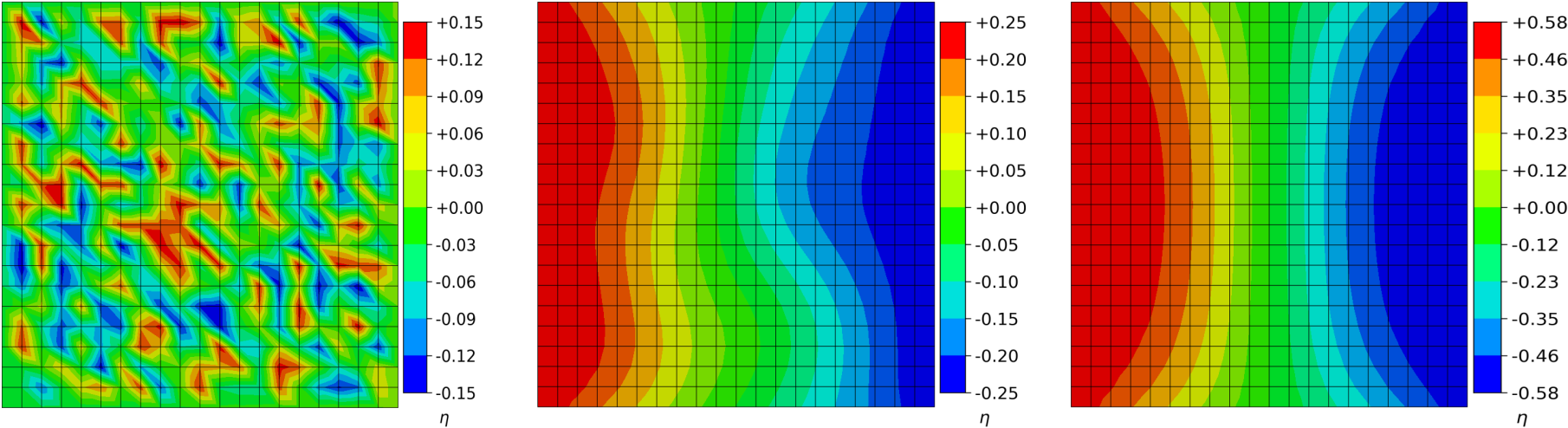}\\
 PBC & \includegraphics[width=\linewidth]{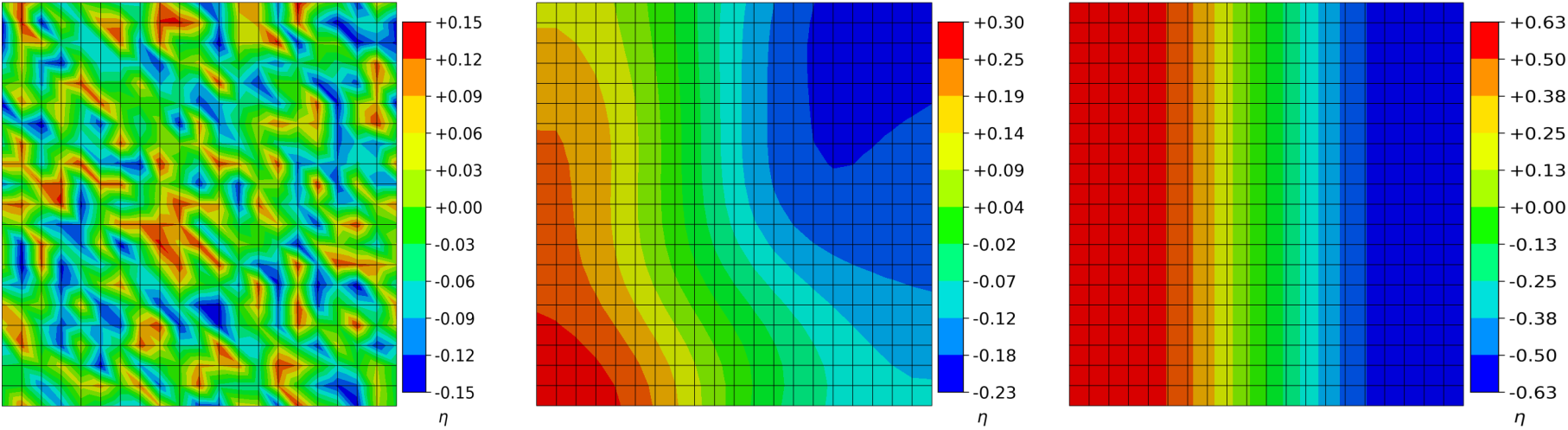}\\
     &  \hspace{1cm}  $t=0$ \hspace{2.7cm}  $t/\intrevoltimescale=0.15$ \hspace{2.5cm}  $t/\intrevoltimescale=32$ \hspace{1cm}
\end{tabular}
\caption{Microscopic phase-field evolution within a single RVE under different boundary conditions: uniform microtraction (UBC), linear order parameter (LBC), and periodic (PBC). A constant macroscopic phase stress of $\phasestressmacro\!=\!0.25\,\wallthickness\,\energybarrier\,\ex$ is prescribed, while the macroscopic order parameter is held at zero ($\orderparammacro\!=\!0$). Snapshots are shown at different normalized times $t/\intrevoltimescale$.}
\label{fig:phaseevolphasestress} 
\end{figure}

\begin{figure}[hbt]
    \centering 
    \hfill
    \begin{subfigure}[b]{0.4\textwidth}
         \centering
         \includegraphics[width=\textwidth]{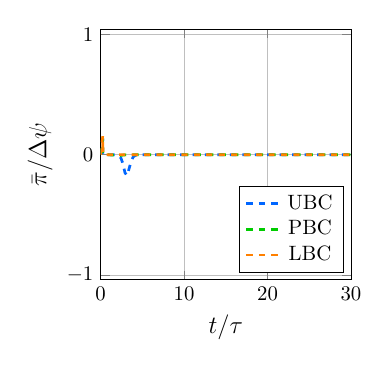}
         \caption{}
         \label{fig:phaseevolphasestressoderparam}
     \end{subfigure}
     \hfill
    \begin{subfigure}[b]{0.4\textwidth}
         \centering
         \includegraphics[width=\textwidth]{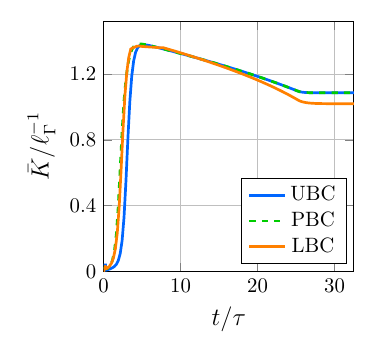}
         \caption{}
         \label{fig:phaseevolphasestressoderparamgrad}
     \end{subfigure}     
     \hfill
\caption{Evolution of the macroscopic microforce, $\microforcemacro$, and gradient of the order parameter, $\orderparamgradmacro$, corresponding to the RVE simulations in Fig.~\ref{fig:phaseevolphasestress}, for a prescribed macroscopic phase stress $\phasestressmacro\!=\!0.25\,\wallthickness\,\energybarrier\,\ex$ under the three different kinds of boundary conditions.}

\label{fig:macroevolphasestress} 
\end{figure}

To understand and verify the predictions of the proposed homogenization theory, we first conduct a series of single RVE simulations under different loading conditions with an RVE size $\RVEedgelength\!=\!1.12\,\wallthickness$. The initial investigation focuses on the influence of the type of boundary conditions applied to the RVE. \figurename~\ref{fig:phaseevolphasestress} illustrates the evolution of the microscopic phase-field within the RVE for three proposed types of boundary conditions: uniform microtraction boundary conditions (UBC), linear order parameter boundary conditions (LBC), and periodic boundary conditions (PBC). 
In this test case, a constant horizontal macroscopic phase stress of $\phasestressmacro=0.25\,\wallthickness\,\energybarrier\,\ex$ is prescribed, while the macroscopic microforce is kept at zero ($\microforcemacro=0$). The corresponding evolution of the macroscopic quantities, namely the order parameter gradient $\orderparamgradmacro$ and the microforce, is depicted in \figurename~\ref{fig:macroevolphasestress}. 
Several key observations can be made. 
First, under all three types of boundary conditions, the system evolves towards a stationary state characterized by a non-zero macroscopic gradient, as expected from the applied loading. However, the microscopic fields differ significantly. For both UBC and PBC, the RVE reaches a stationary state with a domain wall oriented vertically, corresponding to a gradient in the horizontal direction. In contrast, the additional constraint imposed by the LBC introduces a distinct boundary layer at the RVE's top and bottom edges. This is a direct consequence of the incompatibility between the enforced linear profile and the non-linear  solution for a domain wall, which follows a hyperbolic tangent profile due to the non-linearity of the underlying partial differential equation \cite{Steinbach2009}. This microscopic difference has a direct macroscopic consequence. As shown in \figurename~\ref{fig:macroevolphasestress}, the prescribed driving phase stress $\phasestressmacro$ leads to a lower stationary value of the macroscopic gradient $\orderparamgradmacro$ for the LBC case compared to the UBC and PBC cases. This behavior, where the boundary condition type affects the overall response even for a homogeneous material, is a hallmark of non-linear problems and is not observed in standard linear elastic homogenization. It is also worth mentioning that the reorganization of the phase-field at the microscale leads to transient, non-zero values of the macroscopic microforce $\microforcemacro$, which subsequently vanish as the stationary state is approached. 

\subsubsection{Influence of scale separation}
\label{sec:scale_separation}

Classical mechanical homogenization theory, as established by Hill, lacks an intrinsic length scale. Consequently, its predictions are insensitive to a self-similar scaling of the RVE. In contrast, phase-field theory possesses an intrinsic material length scale, $\wallthickness$, as discussed previously. This implies that the size of the RVE, $\RVEedgelength$, relative to this intrinsic length becomes a relevant parameter that governs the macroscopic response.

To investigate this effect, a series of simulations is performed for different ratios of $\wallthickness/\RVEedgelength$. Periodic boundary conditions are applied, and all simulations start from the same interpolated initial order parameter field. The loading is defined by a prescribed macroscopic phase stress $\phasestressmacro=\vect{0}$ and microforce $\microforcemacro=0$.

\begin{figure}[hbtp]
    \centering 
      \begin{tabular}{ m{2.1cm} m{0.76\textwidth} }
    $\frac{\ell_\Gamma}{\ell_{RVE}}=1$ & \includegraphics[width=\linewidth]{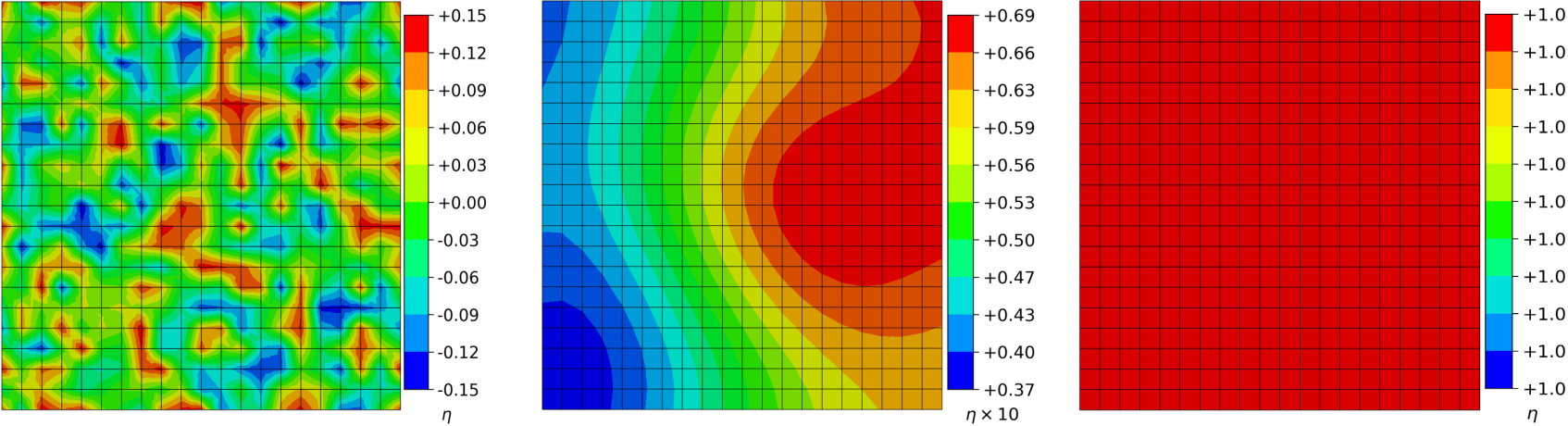}\\ 
    &  \hspace{1cm}  $t=0$ \hspace{2.5cm}  $t/\intrevoltimescale=0.33$ \hspace{2.2cm}  $t/\intrevoltimescale=2.5$ \hspace{1cm} \\[0.05cm]
    $\frac{\ell_\Gamma}{\ell_{RVE}}=0.5$ & \includegraphics[width=\linewidth]{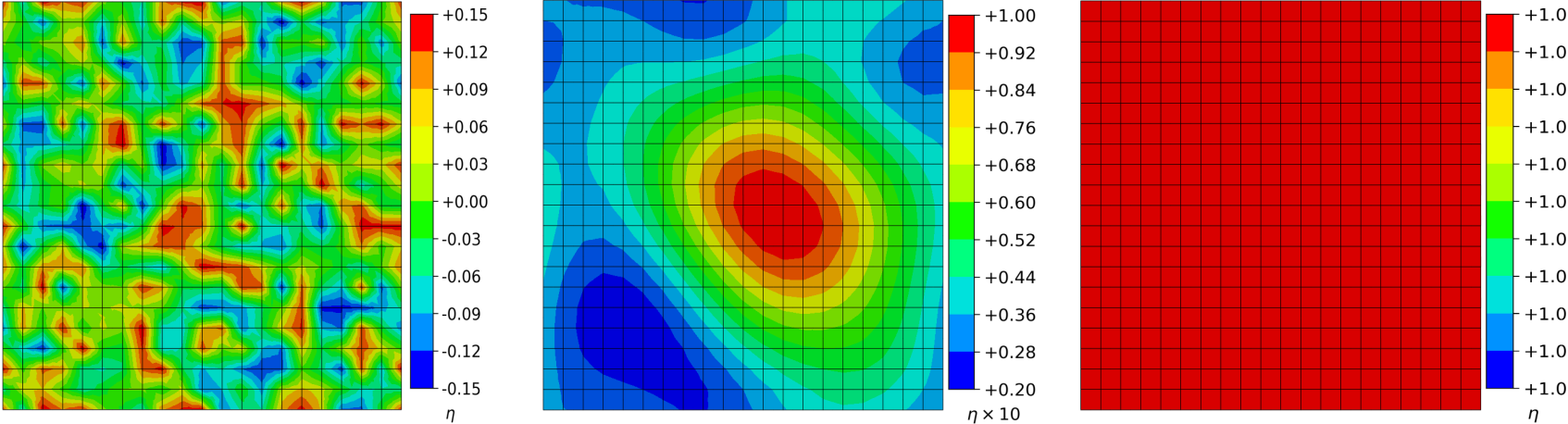}\\
    &  \hspace{1cm}  $t=0$ \hspace{2.5cm}  $t/\intrevoltimescale=0.33$ \hspace{2.2cm}  $t/\intrevoltimescale=5$ \hspace{1cm}\\[0.05cm]
    $\frac{\ell_\Gamma}{\ell_{RVE}}=0.1$ & \includegraphics[width=\linewidth]{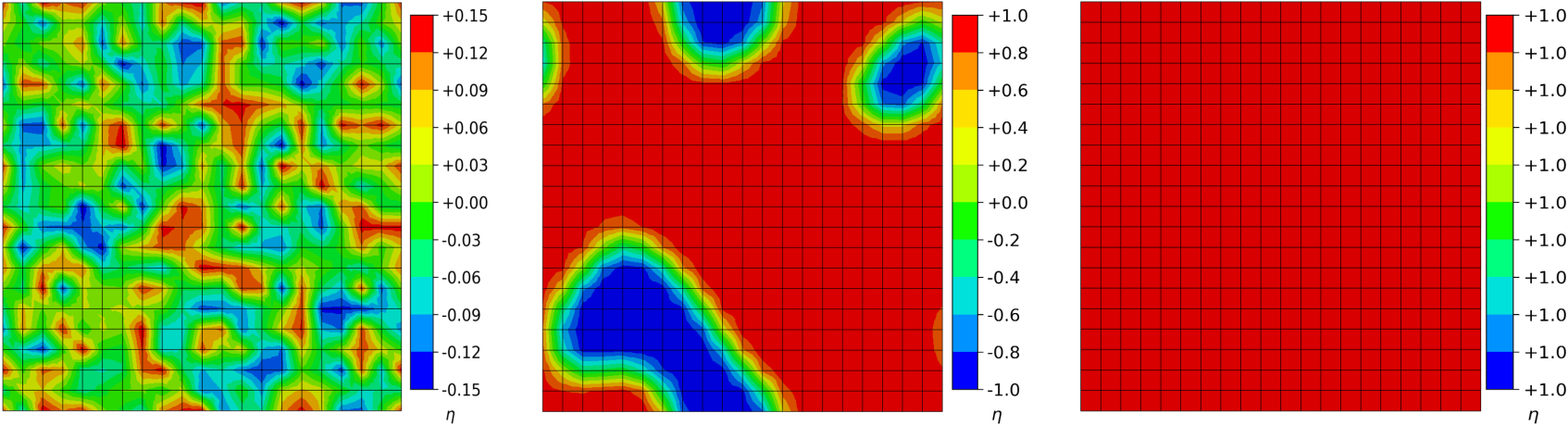}\\
    &  \hspace{1cm}  $t=0$ \hspace{2.5cm}  $t/\intrevoltimescale=1.72$ \hspace{2.2cm}  $t/\intrevoltimescale=14$ 
    \hspace{1cm}\\[0.05cm]
    $\frac{\ell_\Gamma}{\ell_{RVE}}=0.05$ & \includegraphics[width=\linewidth]{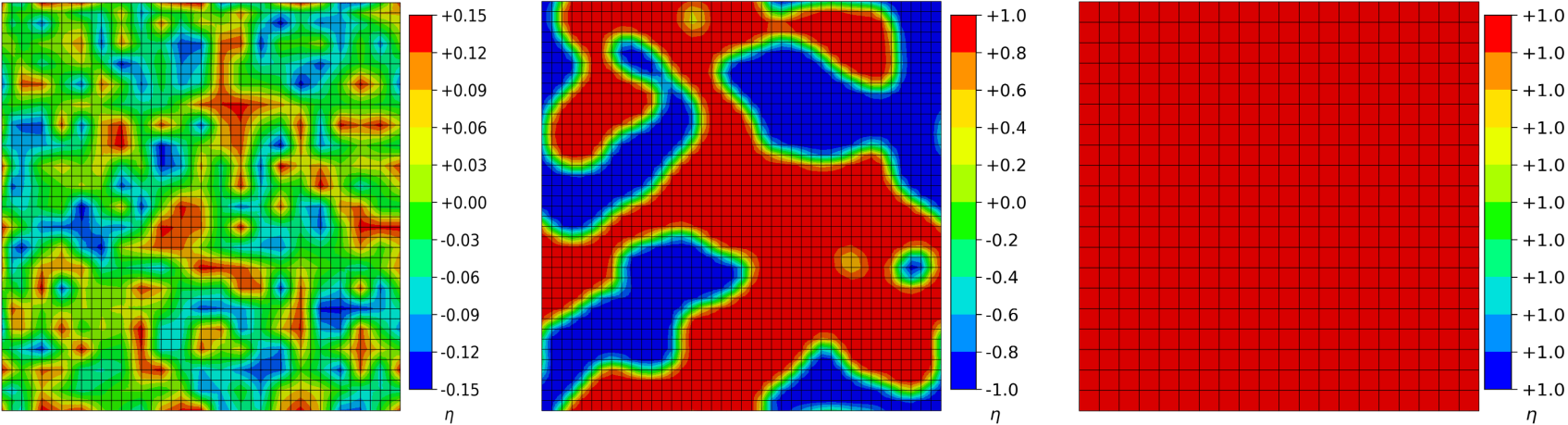}\\
    &  \hspace{1cm}  $t=0$ \hspace{2.5cm}  $t/\intrevoltimescale=1.72$ \hspace{2.2cm}  $t/\intrevoltimescale=56$ \hspace{1cm}\\
  \end{tabular}
    \caption{Snapshots of the microscopic phase evolution within an RVE under periodic boundary conditions, with $\phasestressmacro=\vect{0}$  $\microforcemacro=0$, for different ratios of scale separation, i.e., ratio of wall thickness $\wallthickness$ to RVE edge length $\RVEedgelength$, and suitably refined mesh. All simulations start from the same initial condition and evolve towards a homogeneous state.}
    \label{fig:scalesep_snapshots} 
\end{figure}

Snapshots of the microscopic phase evolution for different scale separation ratios are shown in \figurename~\ref{fig:scalesep_snapshots}. 
As expected, a smaller scale separation (i.e., a larger ratio of $\wallthickness/\RVEedgelength$) results in a smoother, more diffuse phase-field within the RVE. Conversely, a larger scale separation allows for the formation of more complex and fissured intermediate structures during the evolution. Ultimately, however, all simulations reach the same homogeneous stationary state of $\orderparam=1$.

\begin{figure}[hbt]
    \centering 
    \includegraphics[width=0.4\textwidth]{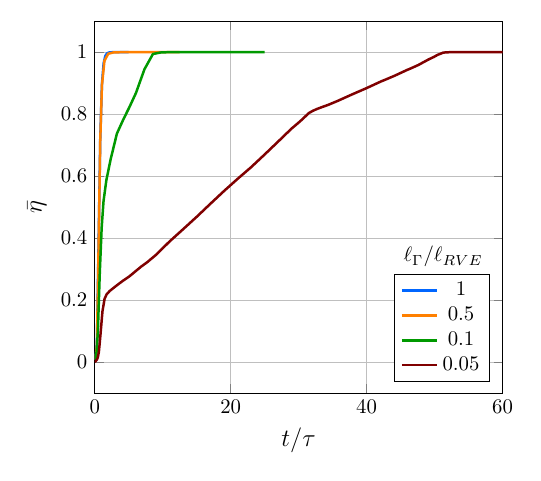}
    \caption{Influence of the scale separation ratio, $\wallthickness/\RVEedgelength$, on the temporal evolution of the macroscopic average phase-field, $\orderparammacro$, at $\phasestressmacro=\vect{0}$ and $\microforcemacro=0$. The results show that a larger separation of scales (smaller ratio) leads to a slower coarsening dynamic and a longer time to reach the stationary state.}
    \label{fig:macroevolscalesep} 
\end{figure}

The evolution of the corresponding macroscopic average order parameter, $\orderparammacro$, is plotted in \figurename~\ref{fig:macroevolscalesep}. 
A clear trend emerges: the larger the scale separation (i.e., the smaller the ratio $\wallthickness/\RVEedgelength$), the longer it takes for the system to reach the final stationary state. This phenomenon, where coarsening dynamics are scale-dependent, is well-known in the broader context of phase-field theory, as discussed for instance by \citet{Bray1994}. On the other end of the spectrum, for low scale separation, the macroscopic response appears to converge as the ratio is increased from $\wallthickness/\RVEedgelength=0.5$ to $1.0$. As seen in \figurename~\ref{fig:scalesep_snapshots}, this regime corresponds to a scenario where the macroscopic response is governed by the growth of a single, almost concave region of the favored phase. In the following, a ratio of $\wallthickness/\RVEedgelength=0.1$ is adopted pragmatically to ensure the presence of complete domain walls within an RVE, while limiting its kinetics to a few domains.

\subsubsection{Notched Structure}
\label{sec:notched_structure}

As an actual application of the proposed FE\textsuperscript{2} framework, we consider a notched structure with a total length of $\ell=79\,\wallthickness$. A notch with a radius of $R=19.75\,\wallthickness$ is introduced to act as a trigger for macroscopic gradients in the phase-field. To enforce an inhomogeneous field distribution, Dirichlet boundary conditions of $\orderparammacro=1$ and $\orderparammacro=-1$ are applied at the top and bottom edges of the structure, respectively, while trivial Neumann boundary conditions are applied to the lateral sides.

For comparison and validation, a direct numerical simulation (DNS) of the entire structure is performed. It is important to note that for a meaningful comparison, the DNS is initialized with a fine periodic pattern, and the same pattern is used for the RVEs in the FE\textsuperscript{2} simulation. To investigate the influence of the initial microstructure, two different random initialization patterns are studied.

\begin{figure}[hbtp]
  \centering
  \begin{subfigure}{\textwidth}
    \includegraphics[width=0.47\linewidth]{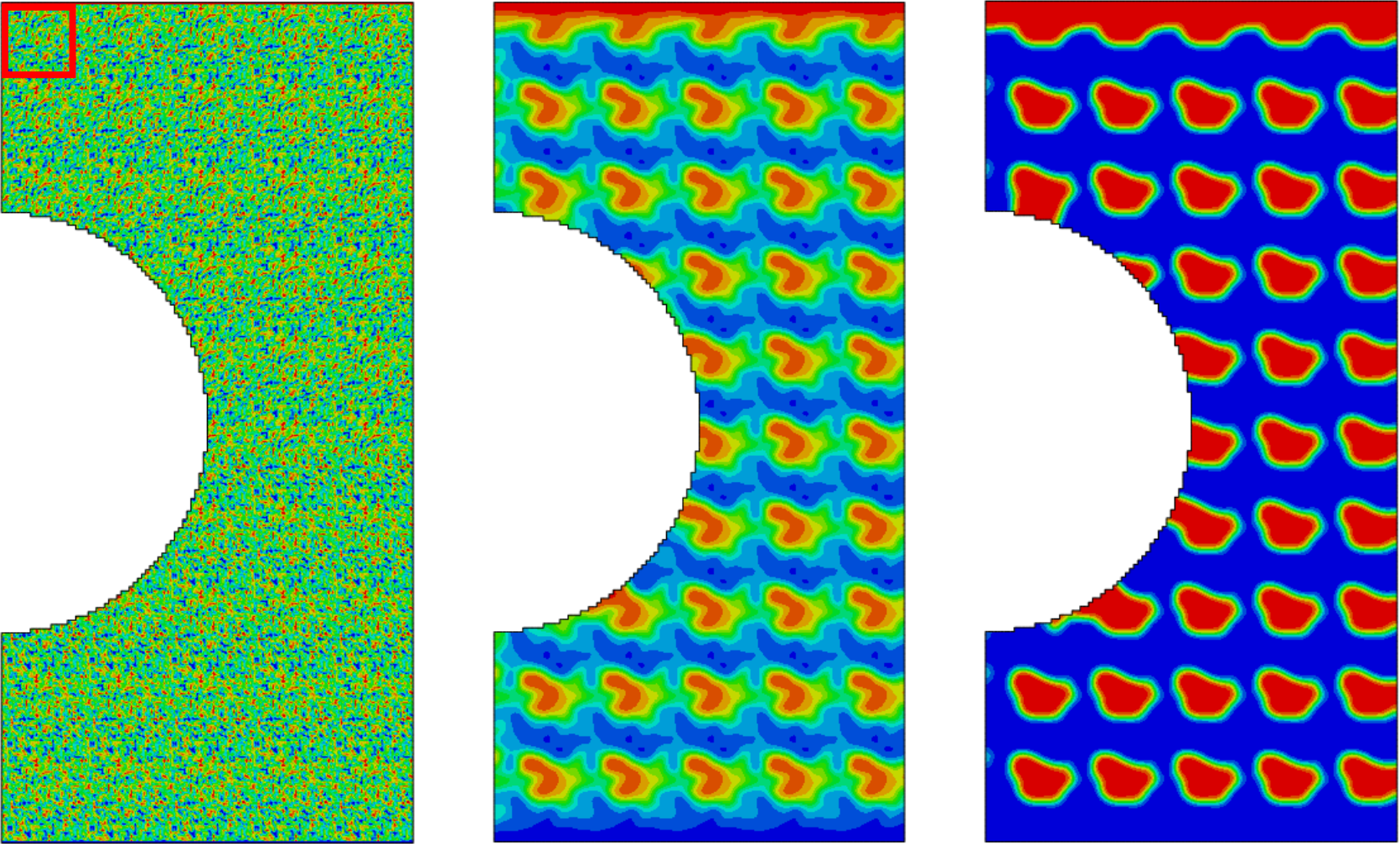}\hspace*{0.3cm}
    \includegraphics[width=0.47\linewidth]{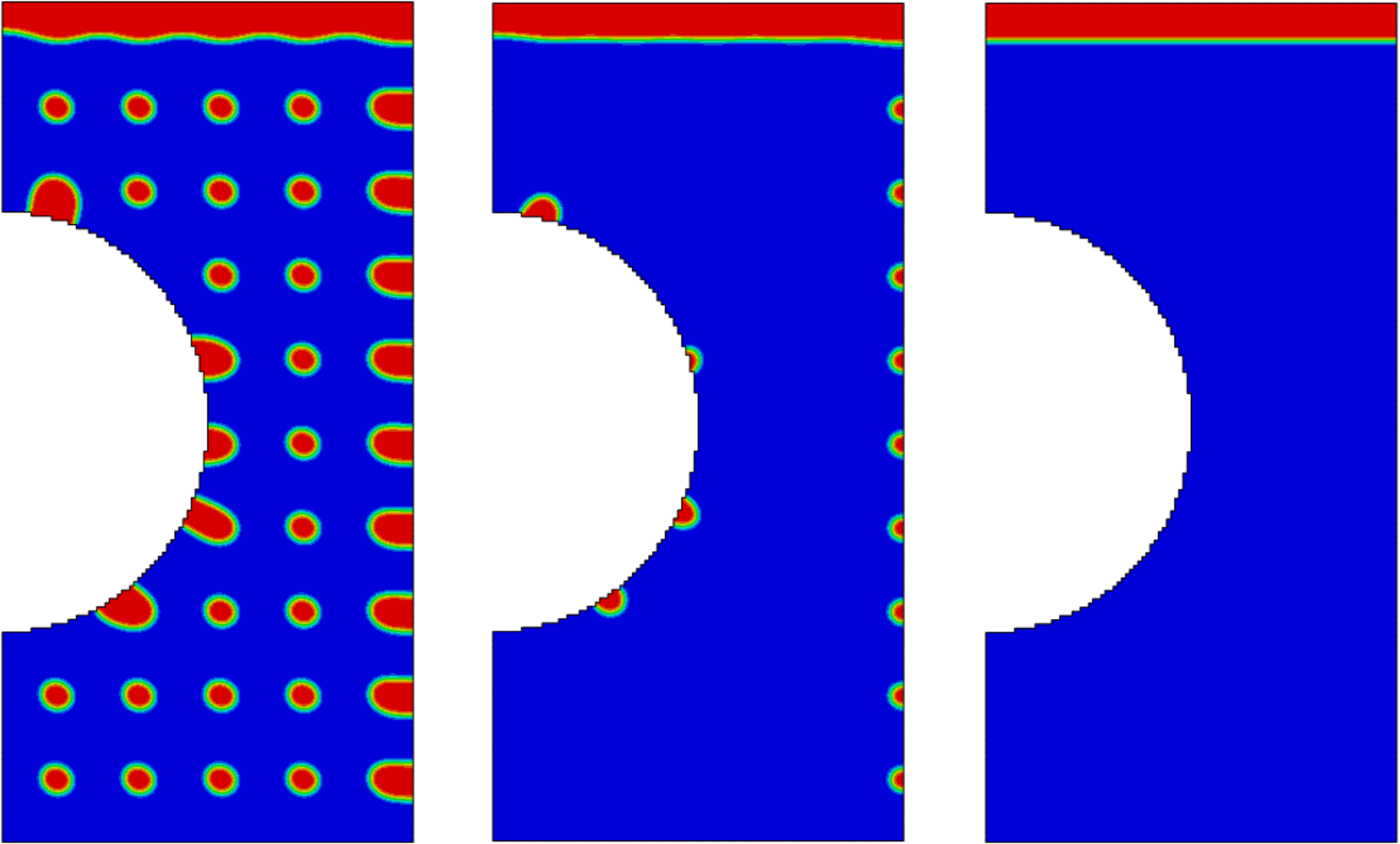}\\
    \hspace*{0.8cm}$t=0$\hfill$t/\intrevoltimescale=0.8$\hfill$t/\intrevoltimescale=1.48$\hfill$t/\intrevoltimescale=10.6$\hfill$t/\intrevoltimescale=21.6$\hfill$t/\intrevoltimescale=428$\hspace*{0.5cm}
    \caption{Direct numerical simulation}
    \label{fig:phasefieldDNS}
  \end{subfigure}
  \begin{subfigure}{\textwidth}
    \vspace*{0.2cm}
    \includegraphics[width=0.47\linewidth]{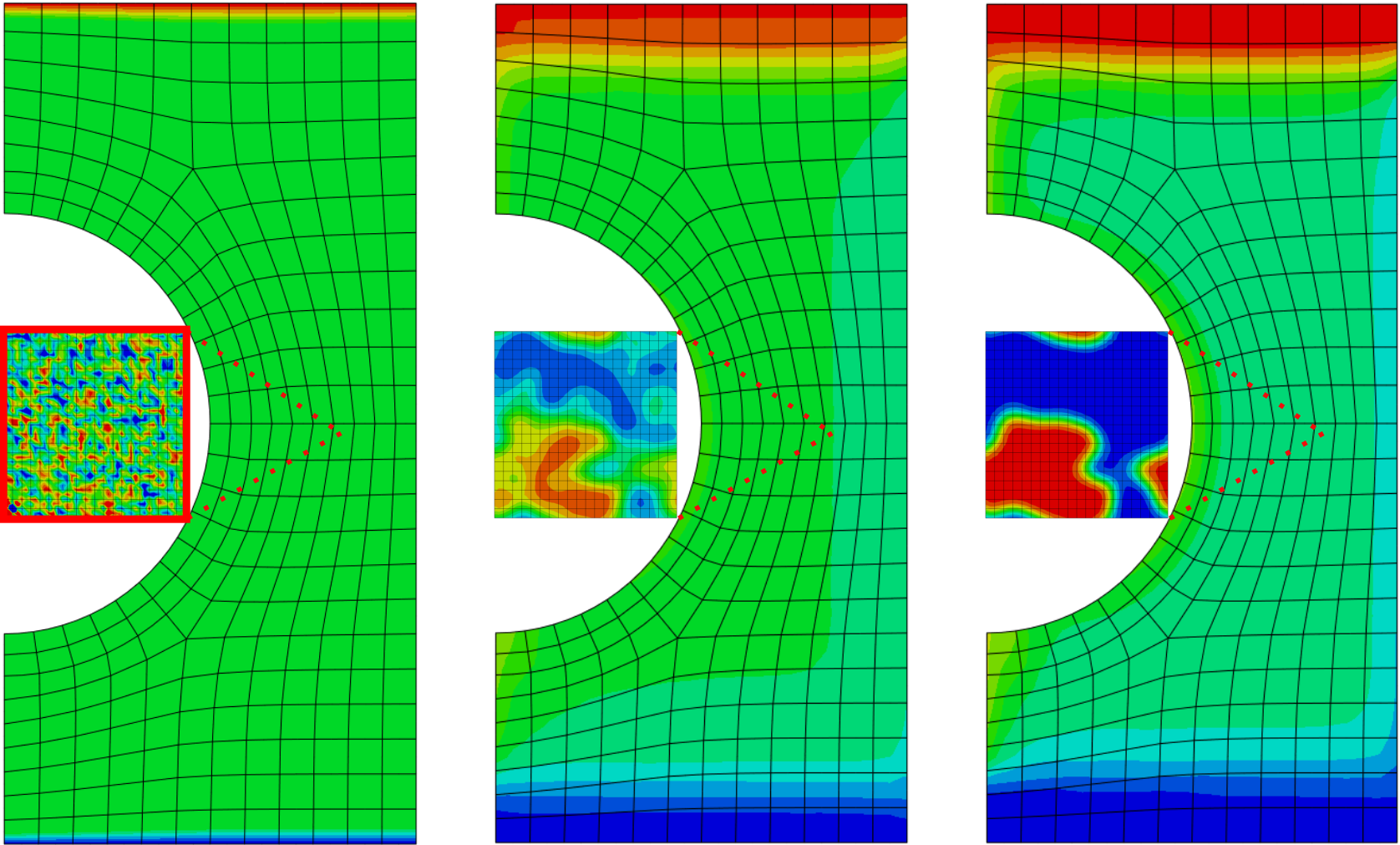}\hspace*{0.3cm}
    \includegraphics[width=0.47\linewidth]{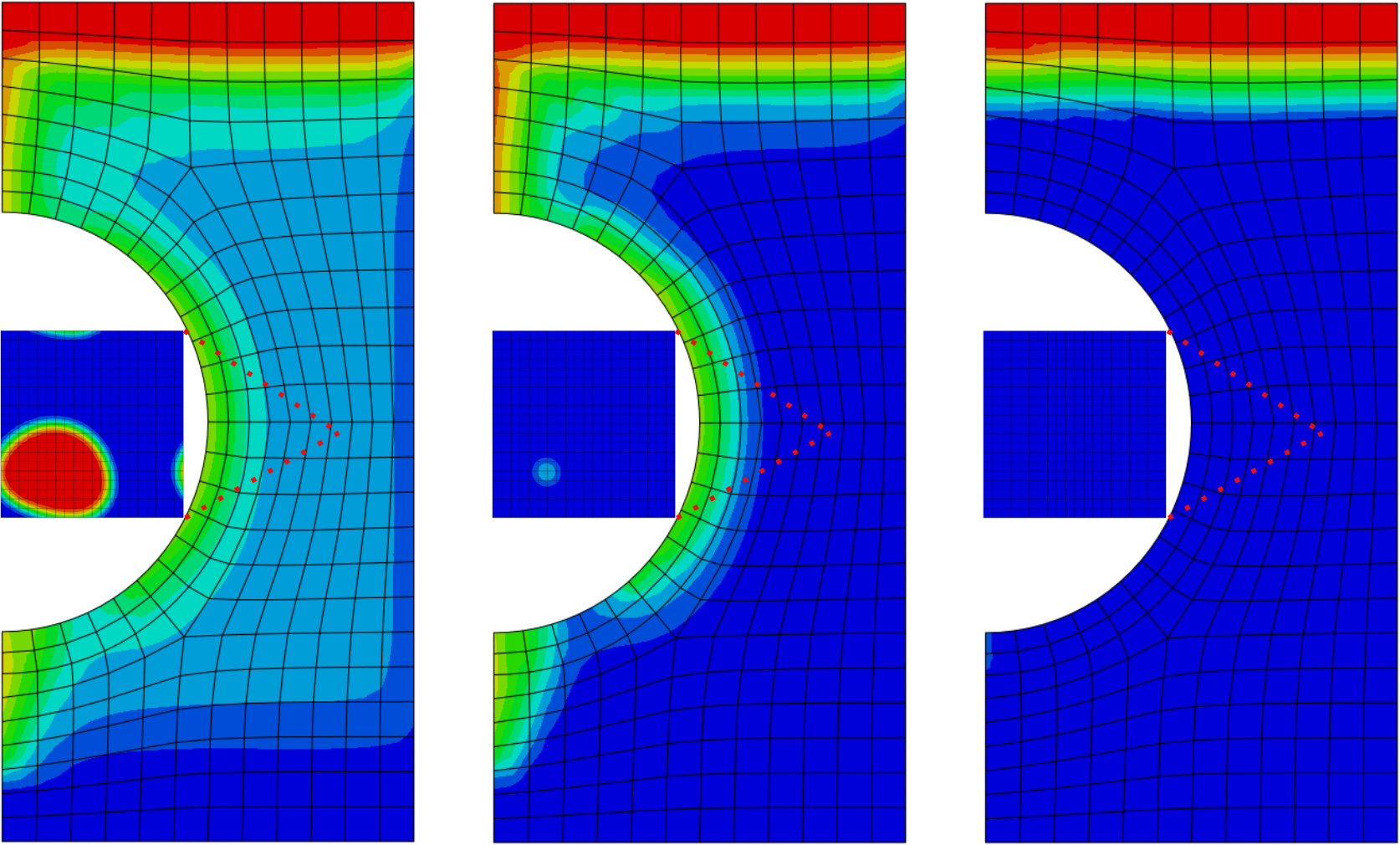}
    \\
    \hspace*{0.8cm}$t=0$\hfill$t/\intrevoltimescale=0.8$\hfill$t/\intrevoltimescale=1.48$\hfill$t/\intrevoltimescale=10.6$\hfill$t/\intrevoltimescale=21.6$\hfill$t/\intrevoltimescale=428$\hspace*{0.5cm}
    \caption{FE\textsuperscript{2} simulation}
    \label{fig:phasefieldFE2}
  \end{subfigure}
  \begin{subfigure}{\textwidth}
    \centering
    \includegraphics[width=0.9\linewidth]{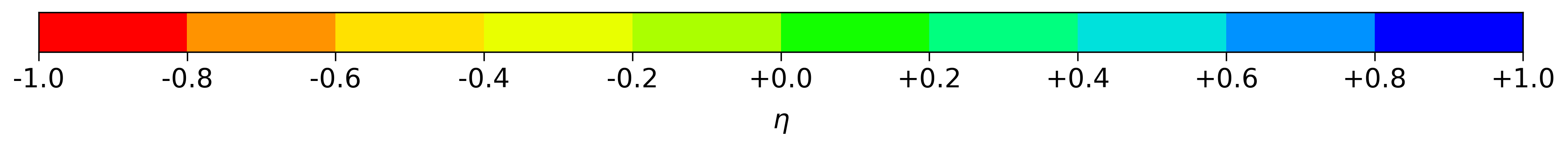}
  \end{subfigure}
   \caption{Comparison of the phase-field evolution in a notched structure for Initialization A. Top row: Direct Numerical Simulation (DNS) with a periodically repeated initial pattern. Bottom row: FE\textsuperscript{2} simulation with the same pattern used in the RVE. The structure is subjected to Dirichlet boundary conditions of $\orderparam=\pm1$ at the top and bottom.}
  \label{fig:phasefieldhole}
\end{figure}

The results of the DNS for the first initial condition (``Initialization A'') are shown in \figurename~\ref{fig:phasefieldDNS}.

As is typical for such Allen-Cahn simulations, the system undergoes a coarsening process where the initial fine structures reorganize into patterns with growing characteristic length scales to minimize the total gradient energy. Due to the applied boundary conditions, the system finally reaches a steady state with a single horizontal domain wall separating the two phases.

The results of the corresponding FE\textsuperscript{2} simulation are presented in \figurename~\ref{fig:phasefieldFE2}. The multiscale model correctly predicts the overall evolution, starting from the random initial field, progressing through intermediate stages with macroscopic \enquote{red islands} of the phase $\orderparam>0$ at the microscale, and ultimately reaching the final state with a single interface. Remarkably, the temporal evolution is also captured with satisfying accuracy.

\begin{figure}[hbt]
  \centering
  \includegraphics[width=0.8\linewidth]{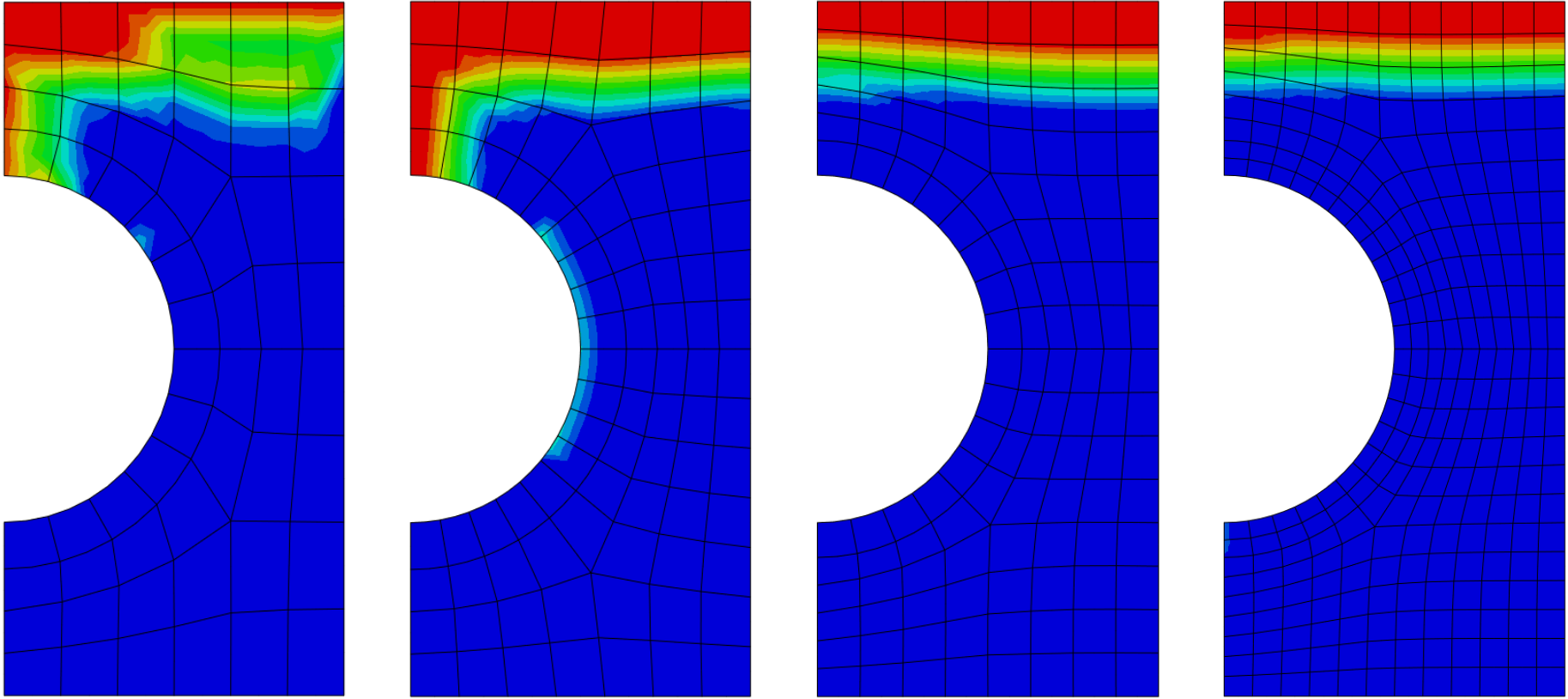}\\
  \includegraphics[width=0.85\linewidth]{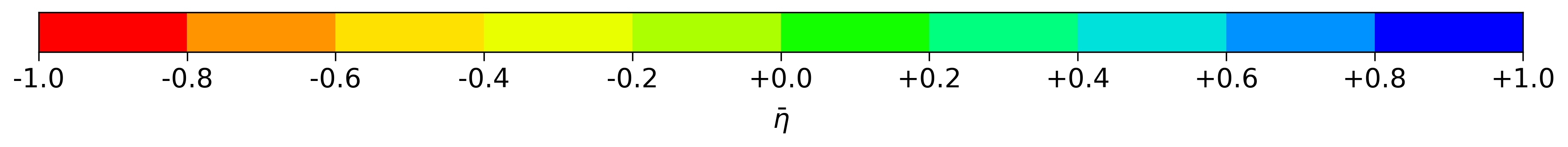}
  \caption[]{Macroscopic mesh convergence study for the macroscopic order parameter at final state $t/\tau=428$ for identical RVE with  Initialization A.}
  \label{fig:phasefieldhole_h-convergence}
\end{figure}

A mesh convergence study was performed with respect to the macroscopic element size, using identical RVEs in terms of size, mesh, and initialization. The corresponding results for the stationary state are presented in Figure~\ref{fig:phasefieldhole_h-convergence}. 
While significant differences are observed between the two coarsest meshes, the final macroscopic order parameter fields for the two finest meshes are virtually identical, indicating convergence within acceptable accuracy requirements. Thus, the mesh (consisting of elements with quadratic shape functions) is sufficiently fine if the transition between the limiting states $\orderparammacro\!=\!-1$ and $\orderparammacro=1$ is resolved by more than a single element.

\begin{figure}[hbtp]
\centering 
\begin{subfigure}{\textwidth}
    \includegraphics[width=0.47\linewidth]{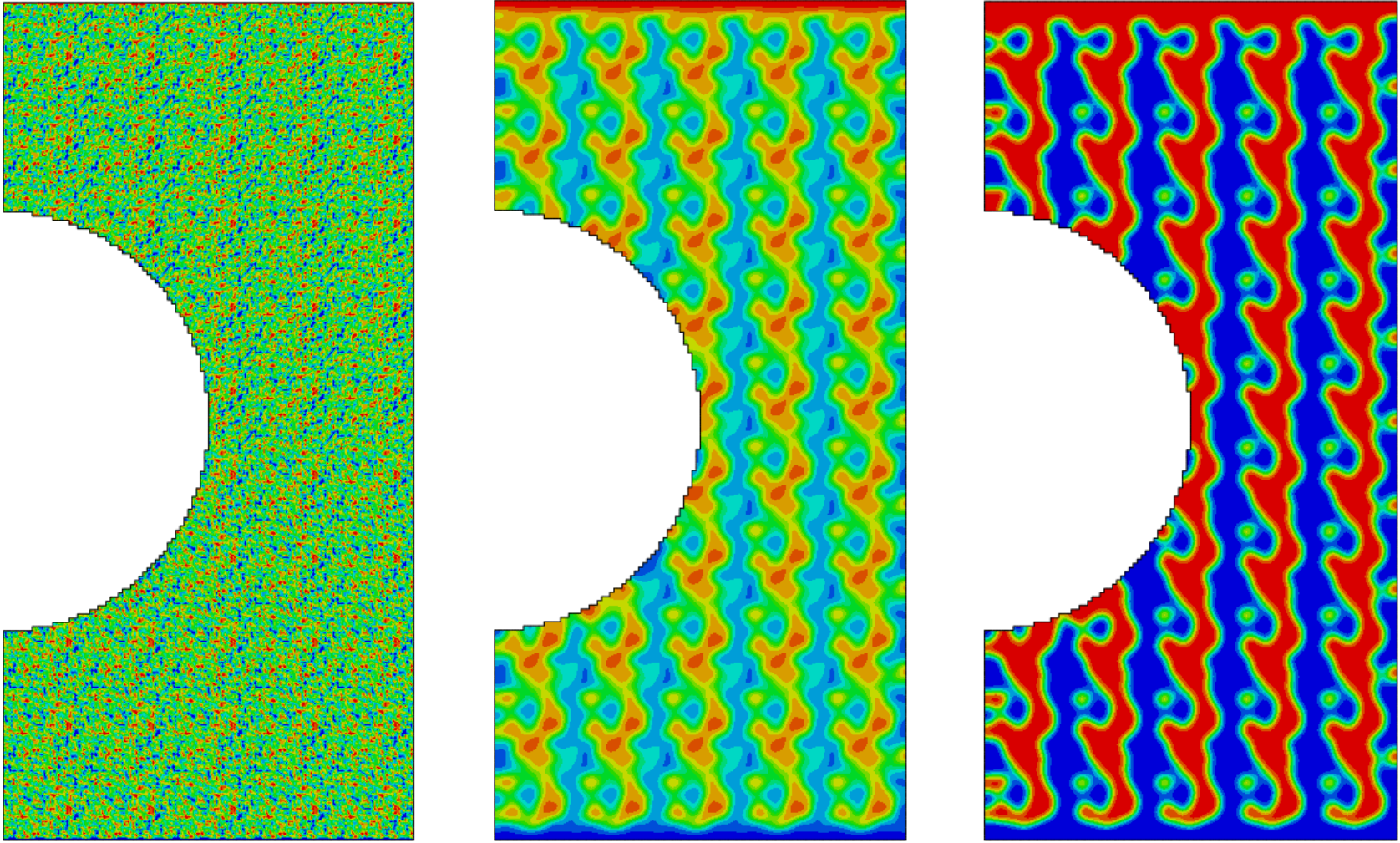}\hspace*{0.3cm}
    \includegraphics[width=0.47\linewidth]{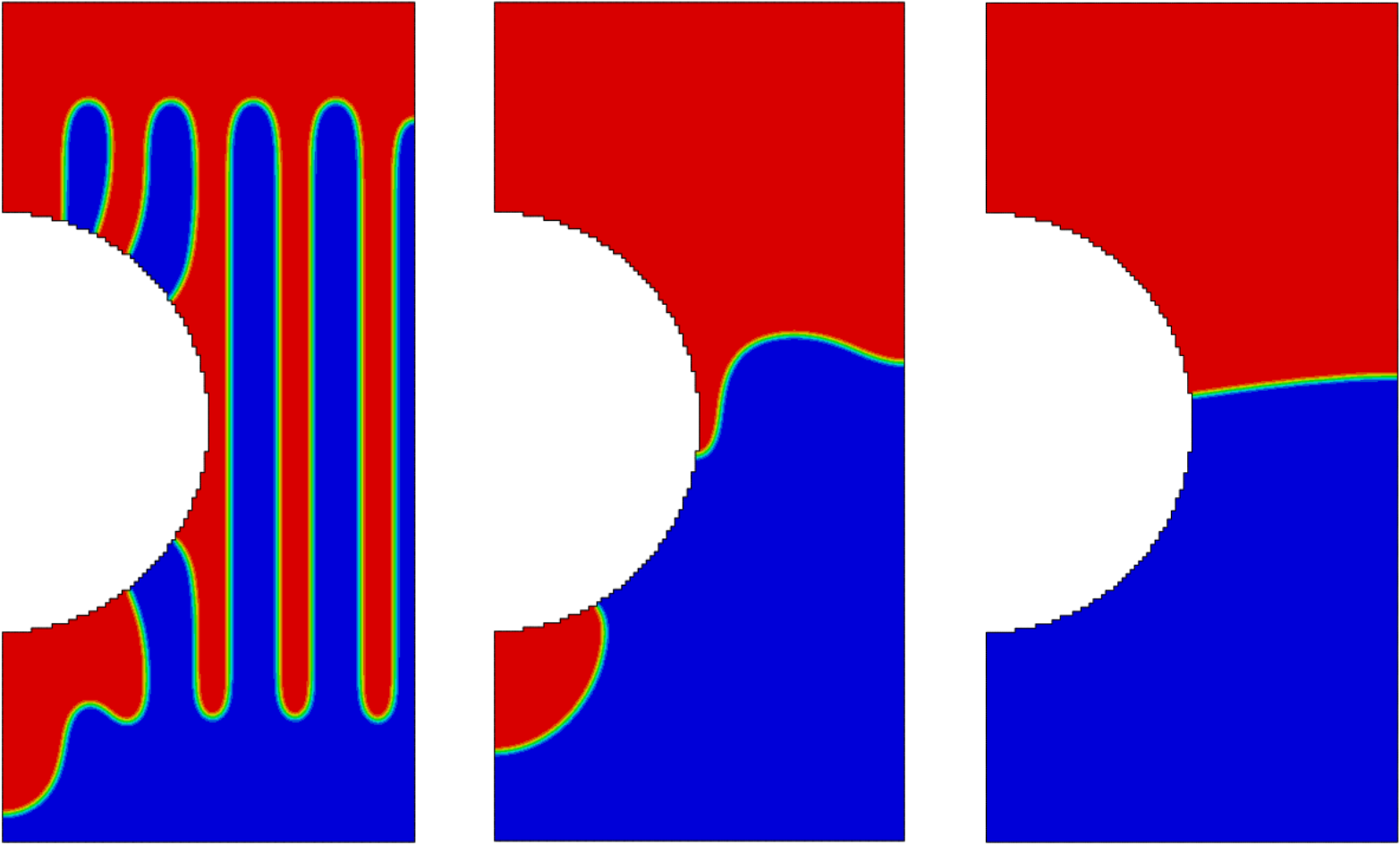}\\
    \hspace*{0.8cm}$t=0$\hfill$t/\intrevoltimescale=0.8$\hfill$t/\intrevoltimescale=1.32$\hfill$t/\intrevoltimescale=41.2$\hfill$t/\intrevoltimescale=188$\hfill$t/\intrevoltimescale=1000$\hspace*{0.4cm}
    \caption{Direct numerical simulation}
    \label{fig:phasefieldDNSB}
\end{subfigure}
\begin{subfigure}{\textwidth}
    \vspace*{0.2cm}
    \includegraphics[width=0.47\linewidth]{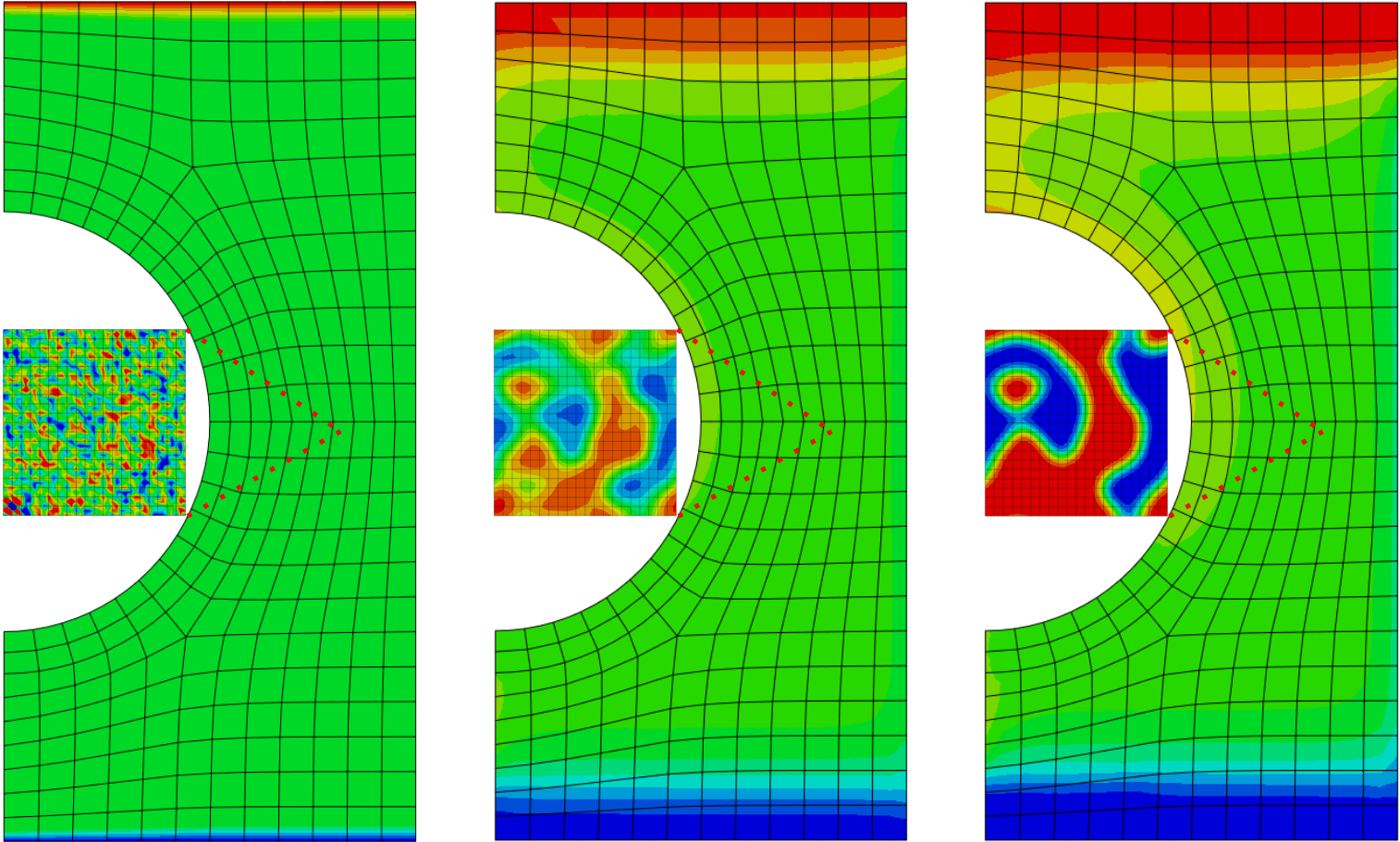}\hspace*{0.3cm}
    \includegraphics[width=0.47\linewidth]{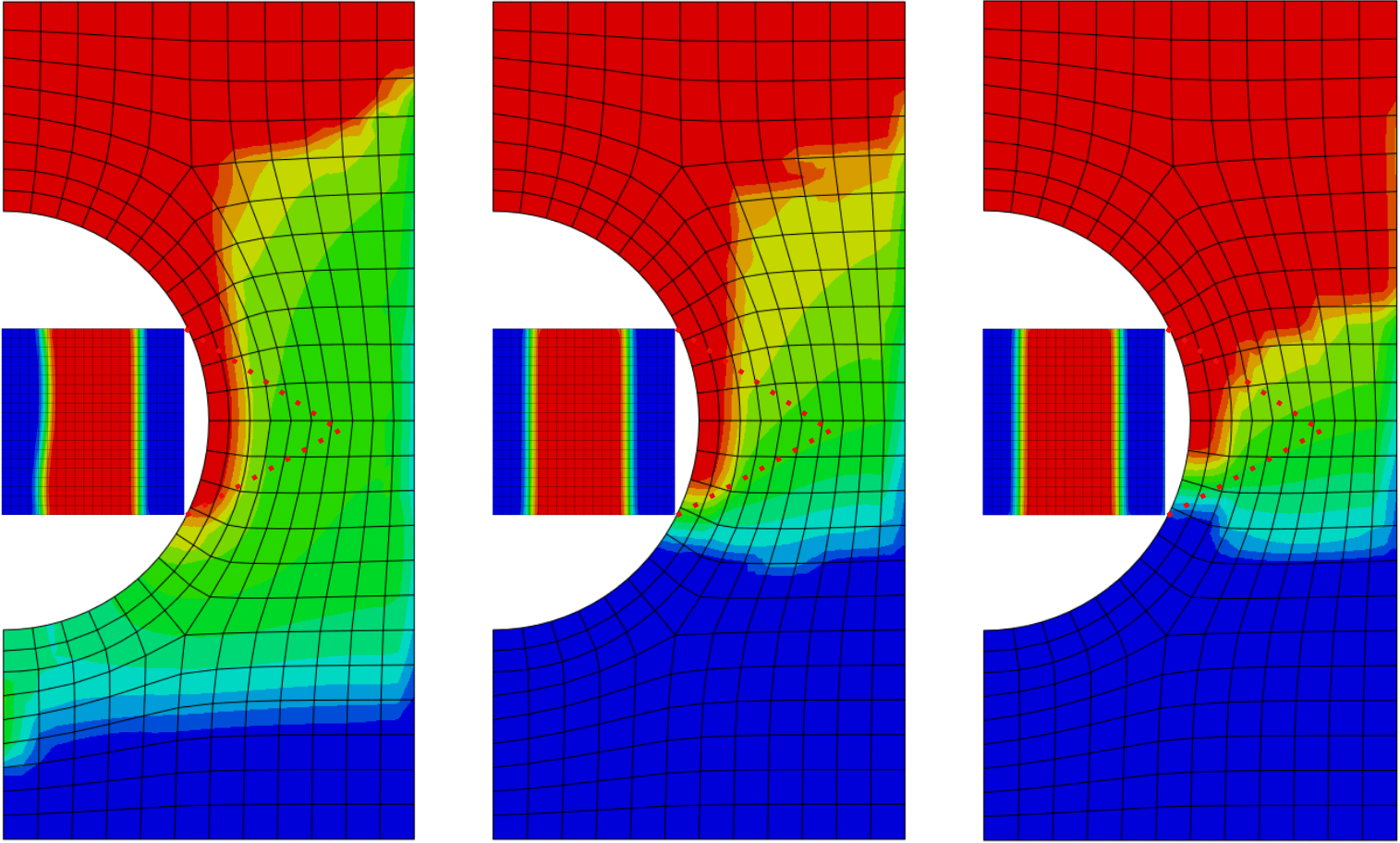}\\
    \hspace*{0.8cm}$t=0$\hfill$t/\intrevoltimescale=0.8$\hfill$t/\intrevoltimescale=1.5$\hfill$t/\intrevoltimescale=41.2$\hfill$t/\intrevoltimescale=188$\hfill$t/\intrevoltimescale=1000$\hspace*{0.4cm}
    \caption{FE\textsuperscript{2} simulation}
    \label{fig:phasefieldFE2B}
\end{subfigure}
\begin{subfigure}{\textwidth}
    \centering
    \includegraphics[width=0.9\linewidth]{figures/Hole_colorbar.png}
  \end{subfigure}
\caption{Comparison of the phase-field evolution in a notched structure for Initialization B. Top row: Direct Numerical Simulation (DNS) with a periodically repeated initial pattern. Bottom row: FE\textsuperscript{2} simulation with the same pattern used in the RVE. The structure is subjected to Dirichlet boundary conditions of $\orderparam=\pm1$ at the top and bottom.}
\label{fig:phasefieldholeB} 
\end{figure}

The simulation is repeated for the same structure, mesh and loading, but with a different random initial field, referred to as ``Initialization B''. The results for both the DNS and the FE\textsuperscript{2} simulation are shown in \figurename~\ref{fig:phasefieldholeB}. Both simulations begin with an initial stage of consolidation, where the random initial field coalesces into larger domains. Subsequently, a notable divergence from the Pattern A simulation occurs: the order parameter field in the DNS evolves towards a distinct pattern of vertical bands in the central region. It is well-known that the Allen-Cahn equation can allow for such layered structures as slowly evolving metastable solutions \cite{Carr1989,Weinberger1985}. Remarkably, the FE\textsuperscript{2} simulation also captures this complex behavior. While the macroscopic order parameter remains close to zero, $\orderparammacro\approx0$, an inspection of the underlying RVEs reveals that the vertical band structure is correctly represented at the microscale. In the final stages of the simulation, these bands vanish completely in the DNS and almost completely in the FE\textsuperscript{2} simulation within the observed time frame, demonstrating good agreement even for these complex, metastable evolution paths.

\subsection{Stress-driven martensitic phase transformation}

\subsubsection{Overview}

To demonstrate the capability of the framework to handle mechanically-coupled problems, we now consider the stress-driven martensitic phase transformation. As a representative example, we adopt the model presented by \citet{Rajendran2020} for the transformation from a parent austenite phase, represented by $\orderparam=0$, into two distinct martensitic product phases, or variants, represented by $\orderparam=\pm1$.

Without going into full detail, the essential aspects of the theory shall be briefly recalled here. The free energy is composed of a chemical contribution, $\Helmenergy_\mathrm{ch}$, and a mechanical contribution, $\Helmenergy_\mathrm{m}$:
\begin{align}
    \Helmenergy_\mathrm{ch}(\orderparam,\temperature)&=\frac12 \,A(\temperature)\,\orderparam^2+\left[3\,\Delta G(\temperature)-A(\temperature)\right]\orderparam^4+\frac{1}{2}\left[A(\temperature)-4\,\Delta G(\temperature)\right]\orderparam^6\ ,\\[.5ex]
    \Helmenergy_\mathrm{m}(\strain,\orderparam) &= \frac12 \left[\strain-\straintr(\orderparam)\right]\ssprod\tstiff(\orderparam)\ssprod\left[\strain-\straintr(\orderparam)\right]
\end{align}
The Allen-Cahn type dissipation potential from Eq.~\eqref{eq:DissPotAllenCahn} and the isotropic gradient energy term from Eq.~\eqref{eq:freeenergyAllenCahn} remain unchanged.

The chemical free energy, $\Helmenergy_\mathrm{ch}$, is a sixth-order polynomial in the order parameter $\orderparam$. This specific form creates a metastable energy landscape that allows the phase transformation to be driven by mechanical loading. This is particularly effective in the temperature range between the equilibrium temperature $\temperature_\mathrm{e}$ (where austenite and martensite have equal free energy) and a critical temperature $\temperature_\mathrm{c}$ (below which austenite becomes unstable). The model assumes a linear temperature dependency for the chemical driving forces via the relations $\Delta G(T)\!=\!G_0\, [\temperature-\temperature_\mathrm{e}]$ and $A(\temperature)\!=\!A_0\,[\temperature-\temperature_\mathrm{c}]$.

The mechanical coupling is introduced through the transformation strain, $\straintr(\orderparam)$, which represents the lattice distortion during the phase change:
\begin{equation}
    \straintr(\orderparam)=\varphi(\orderparam)\left[\,\straintr[\mathrm{p}]+\mathrm{sgn}(\eta)\,\straintr[\mathrm{v}] \,\right]\ .
\end{equation}
Here, $\straintr[\mathrm{p}]$ is the strain associated with the transformation from austenite to martensite, while $\straintr[\mathrm{v}]$ accounts for the transformation strain between the two martensitic variants. The interpolation function $\varphi(\orderparam)\!=\!\frac12\, \orderparam^2\left[3-\orderparam^4\right]$ ensures a smooth transition. The same function is used to interpolate the stiffness tensor between the austenite stiffness $\tstiff_0$ and the martensite stiffness $\tstiff_1$:
\begin{equation}
    \tstiff(\orderparam)=\left[1-\varphi(\orderparam)\right]\tstiff_0+\varphi(\orderparam)\,\tstiff_1
\end{equation}
The material parameters for zirconia are adopted from \citet{Rajendran2020}: $\temperature_\mathrm{e}\!=\!\SI{1305}{\kelvin}$, $\temperature_\mathrm{c}\!=\!\SI{1367}{\kelvin}$, $A_0\!=\!\SI{5.53164e-4}{\giga\pascal\per\kelvin}$, and $G_0\!=\!\SI{1.84388e-4}{\giga\pascal\per\kelvin}$. For the specific values of the transformation strain and stiffness tensors, the reader is referred to the original work \cite{Rajendran2020}. All simulations are performed isothermally at a temperature of \temperature=\SI{1050}{\kelvin}, where austenite is stable under stress-free conditions, but the energy barrier is low enough that mechanical stresses can drive the transformation towards martensite (TRIP effect).

The numerical implementation again utilizes the heat transfer analogy, but now employs coupled thermo-mechanical elements to account for the mechanical fields. Specifically, four-node quadrilateral plane strain elements with temperature degrees of freedom (Abaqus internal name: \texttt{CPE4T}) are used. It is important to keep in mind that the thermal degree of freedom in these finite element simulations corresponds to the order parameter $\orderparam$, while the actual temperature is fixed as described above.

\subsubsection{Single RVE simulation}

Also for the mechanically-coupled case, the analysis begins with a single RVE simulation to probe the fundamental response of the homogenized material.

\begin{figure}[hbt]
    \centering 
     \includegraphics[width=0.48\linewidth]{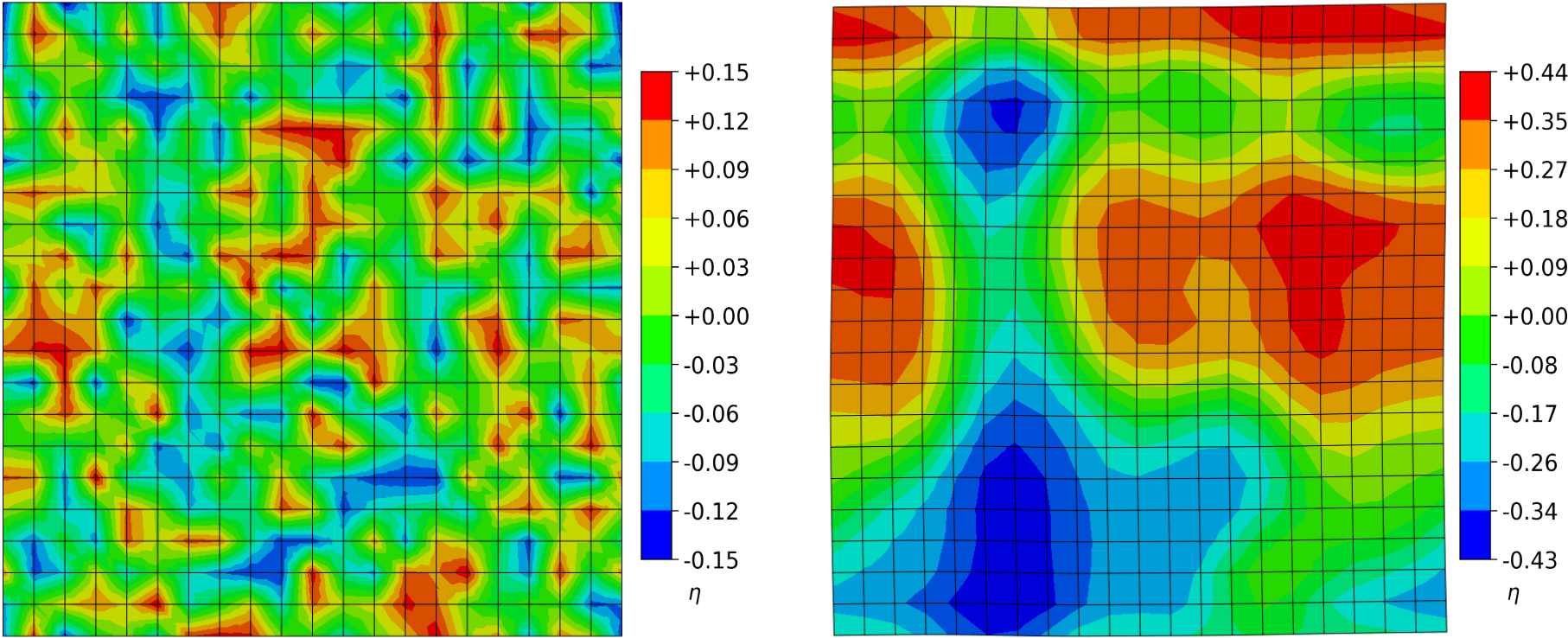}\hfill
     \includegraphics[width=0.48\linewidth]{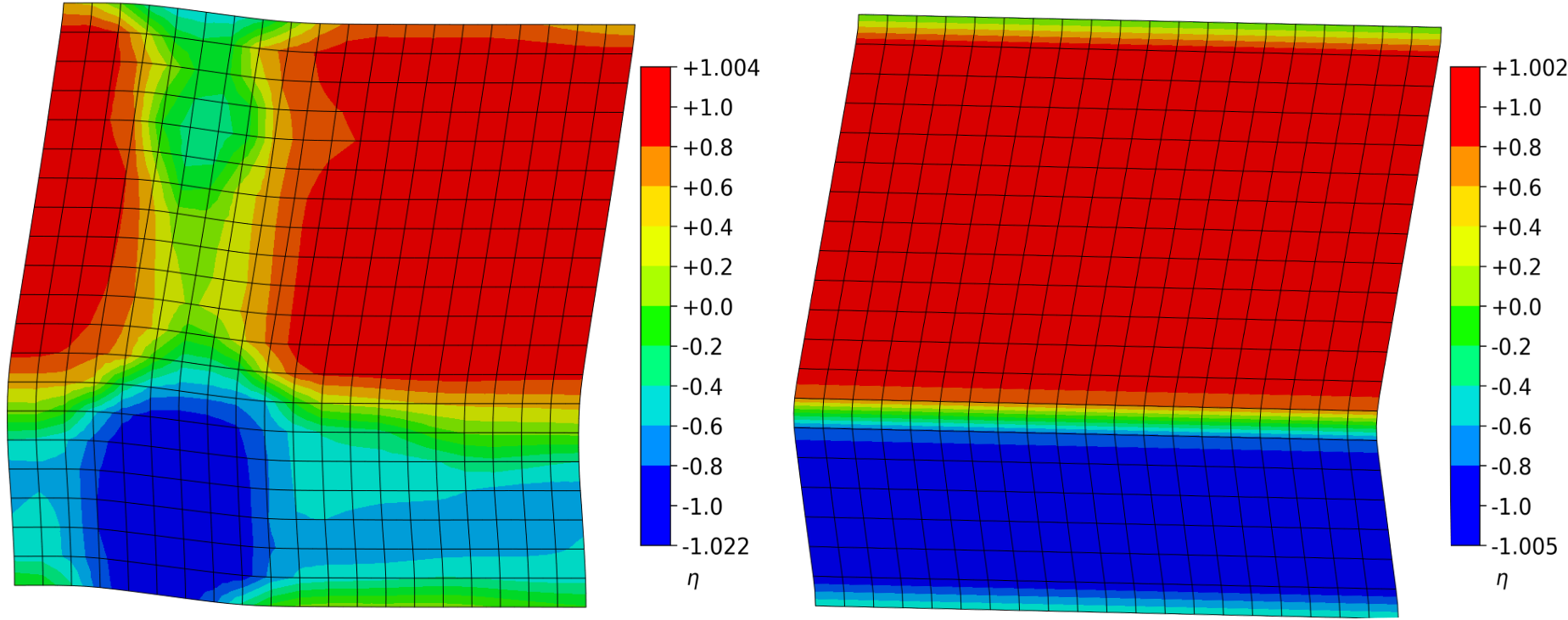}\\
     \hspace{1cm}  $t=0$ \hfill  $t/\intrevoltimescale=0.96$ \hfill  $t/\intrevoltimescale=2.1$ \hfill  $t/\intrevoltimescale=8$ \hspace{1cm}
    \caption[]{Evolution of the martensitic microstructure within a single RVE for the mechanically-coupled model under a macroscopic order parameter gradient of $\orderparamgradmacrocomp_y\!=\!0.05\,\wallthickness^{-1}$ and zero macroscopic stress and microforce ($\stressmacro\!=\!0$, $\microforcemacro\!=\!0$). The system evolves from a random initial state to a twinned structure.
}
    \label{fig:mechanicalRVE_snapshots} 
\end{figure}

The focus of this investigation is placed on the effect of the macroscopic gradient of the order parameter, $\orderparamgradmacro$, as the consistent handling of this quantity is a main novel contribution of the proposed theory.
\figurename~\ref{fig:mechanicalRVE_snapshots} shows the evolution of the phase-field at the microscale. A ramp-wise vertical gradient of the macroscopic order parameter, $\orderparamgradmacro$, is prescribed, while the macroscopic stress and microforce are held at zero. The RVE evolves from a random initial state into a finely twinned martensitic structure. 

\begin{figure}[hbt]
    \centering 
     \hfill
    \begin{subfigure}[b]{0.32\textwidth}
         \centering
         \includegraphics[width=\textwidth]{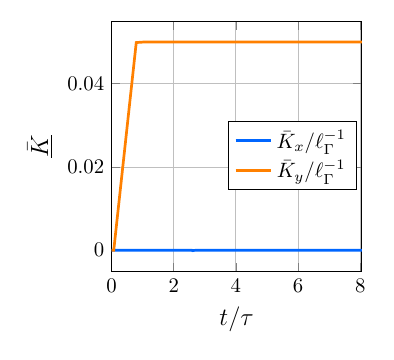}
         \caption{}
         \label{fig:macroevolmechphasegradoderparamgrad}
     \end{subfigure}   
     \hfill
     \begin{subfigure}[b]{0.32\textwidth}
         \centering
         \includegraphics[width=\textwidth]{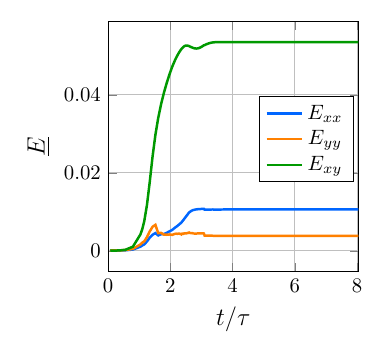}
         \caption{}
         \label{fig:macroevolmechphasegradmech}
     \end{subfigure}         
    \hfill
    \begin{subfigure}[b]{0.32\textwidth}
         \centering
         \includegraphics[width=\textwidth]{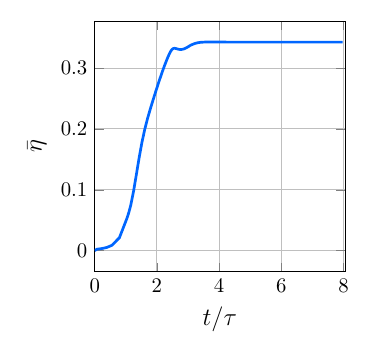}
         \caption{}
         \label{fig:macroevolmechphasegradoderparam}
     \end{subfigure}
 
\caption[]{Evolution of the macroscopic average phase-field, $\orderparammacro$, its gradient $\orderparamgradmacro$ and strain, $\strainmacro$, for the mechanically-coupled RVE simulation shown in Fig.~\ref{fig:mechanicalRVE_snapshots} for a constant prescribed macroscopic order parameter gradient of $\orderparamgradmacrocomp_y\!=\!0.05\,\wallthickness^{-1}$.}
\label{fig:macroevolmechphasegrad} 
\end{figure}

The evolution of the macroscopic quantities is shown in \figurename~\ref{fig:macroevolmechphasegrad}. \figurename~\ref{fig:macroevolmechphasegradoderparamgrad} simply depicts the prescribed loading history of the macroscopic order parameter gradient. More interestingly, since neither a macroscopic microforce ($\microforcemacro\!=\!0$) nor a macroscopic stress ($\stressmacro=0$) was prescribed, the system was not constrained to a symmetric state with a zero average order parameter. Instead, as shown in \figurename~\ref{fig:macroevolmechphasegradmech}, the RVE evolves to a state with a non-zero average order parameter, $\orderparammacro$. Due to the chemo-mechanical coupling, this results in the emergence of a corresponding non-zero macroscopic transformation strain, $\strainmacro$.

\subsubsection{FE\textsuperscript{2} simulation of a plate with a hole}

\begin{figure}
    \centering
    \includegraphics[width=0.3\linewidth]{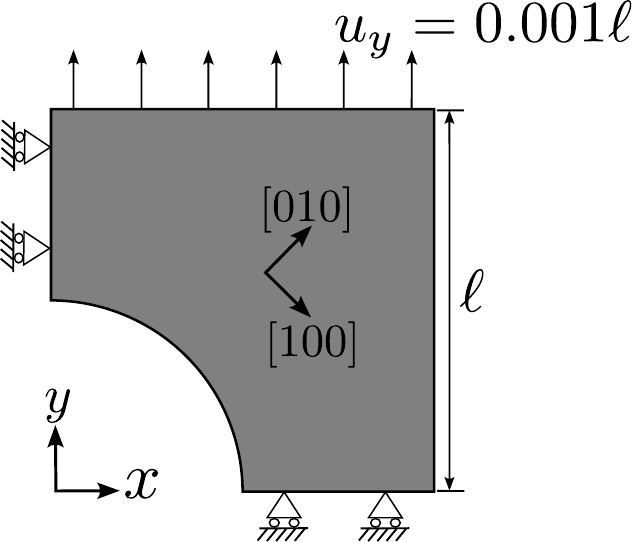}
    \caption{Quarter model of a plate with a hole under tension with crystallographic axes aligned by 45\textdegree{} against the direction of loading.}
    \label{fig:platewithhole}
\end{figure}

The final example considers a plate with a central hole, which serves as a common benchmark problem for generating inhomogeneous fields. A quarter model of the plate is used, with a hole radius of $0.5\,\ell$ and a plate width of $\ell=269\,\wallthickness$, as depicted in \figurename~\ref{fig:platewithhole}. 
The plate is subjected to uniaxial tension via a prescribed displacement $u_y$ at the top edge, which is ramped up linearly until $t\!=\!2.9\,\intrevoltimescale$. For validation, a direct numerical simulation (DNS) is performed with a fine mesh resolution of $h=0.001\,\ell$. Crucially, both the DNS and the RVEs in the multiscale model are initialized with the same periodic random pattern again. It should also be noted that the principal crystallographic axes of the material are rotated by \SI{45}{\degree} with respect to the global $x$-$y$ coordinate system shown.

\begin{figure}[hbtp]
    \centering 
    \begin{subfigure}[b]{\textwidth}
        \centering
        \begin{subfigure}[b]{0.3\textwidth}
            \centering
            \includegraphics[width=\linewidth]{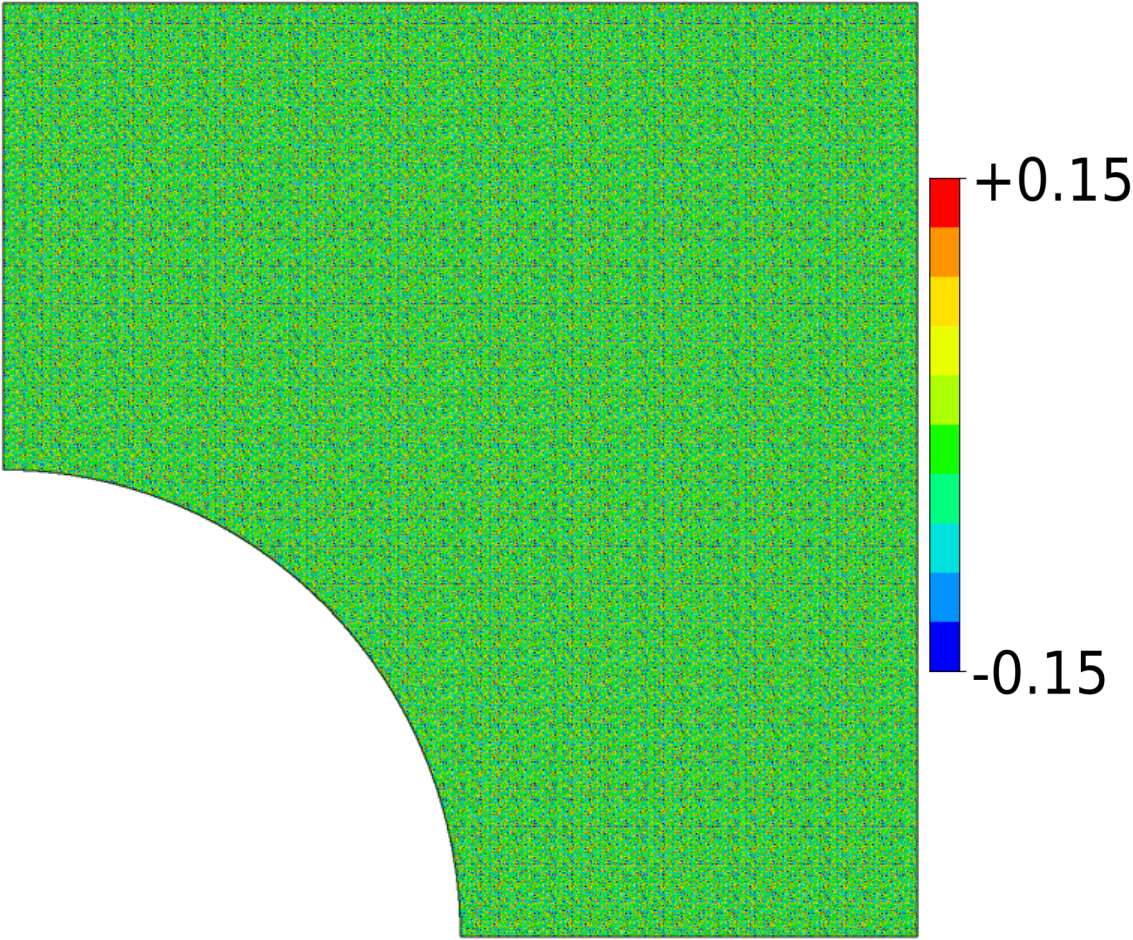}
            {\small $t=0$}
        \end{subfigure}
        \hfill
        \begin{subfigure}[b]{0.3\textwidth}
            \centering
            \includegraphics[width=\linewidth]{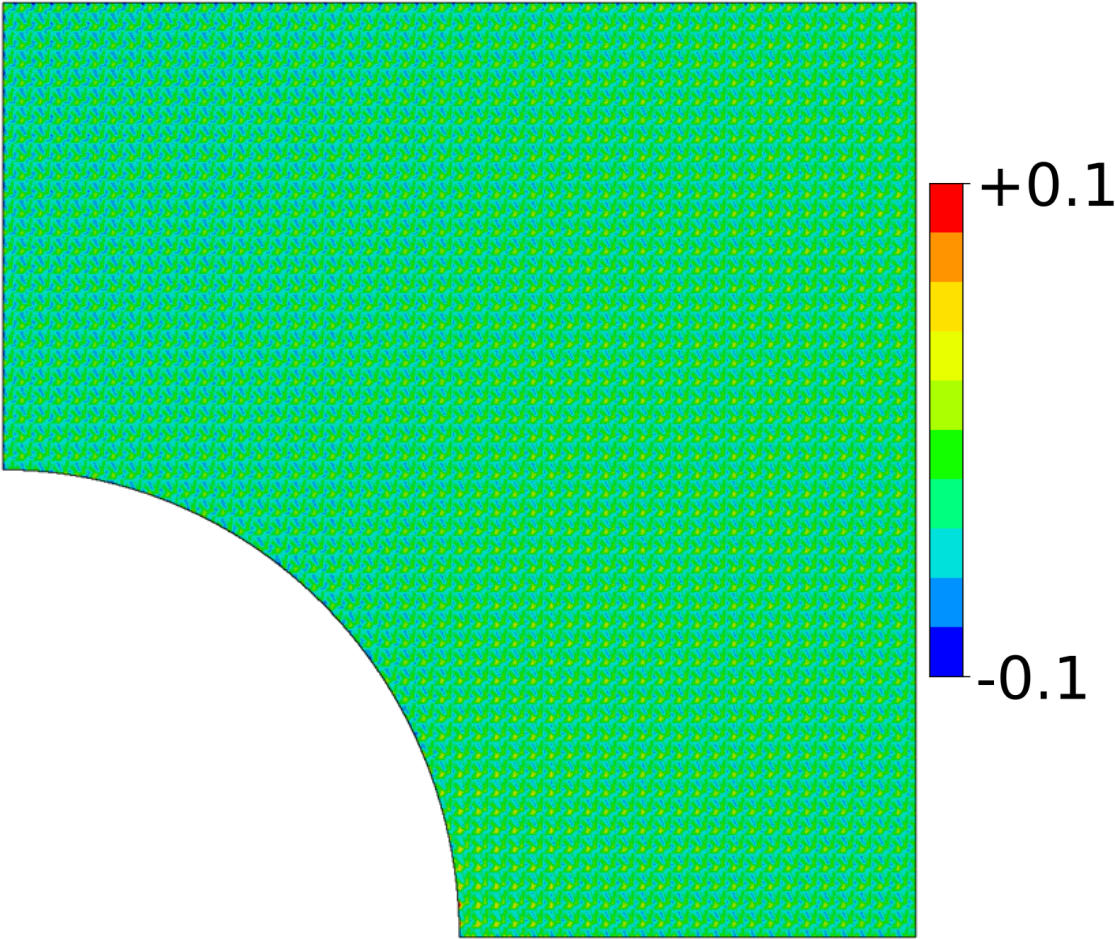}
            {\small $t/\tau=0.6$}
        \end{subfigure}
        \hfill
        \begin{subfigure}[b]{0.3\textwidth}
            \centering
            \includegraphics[width=\linewidth]{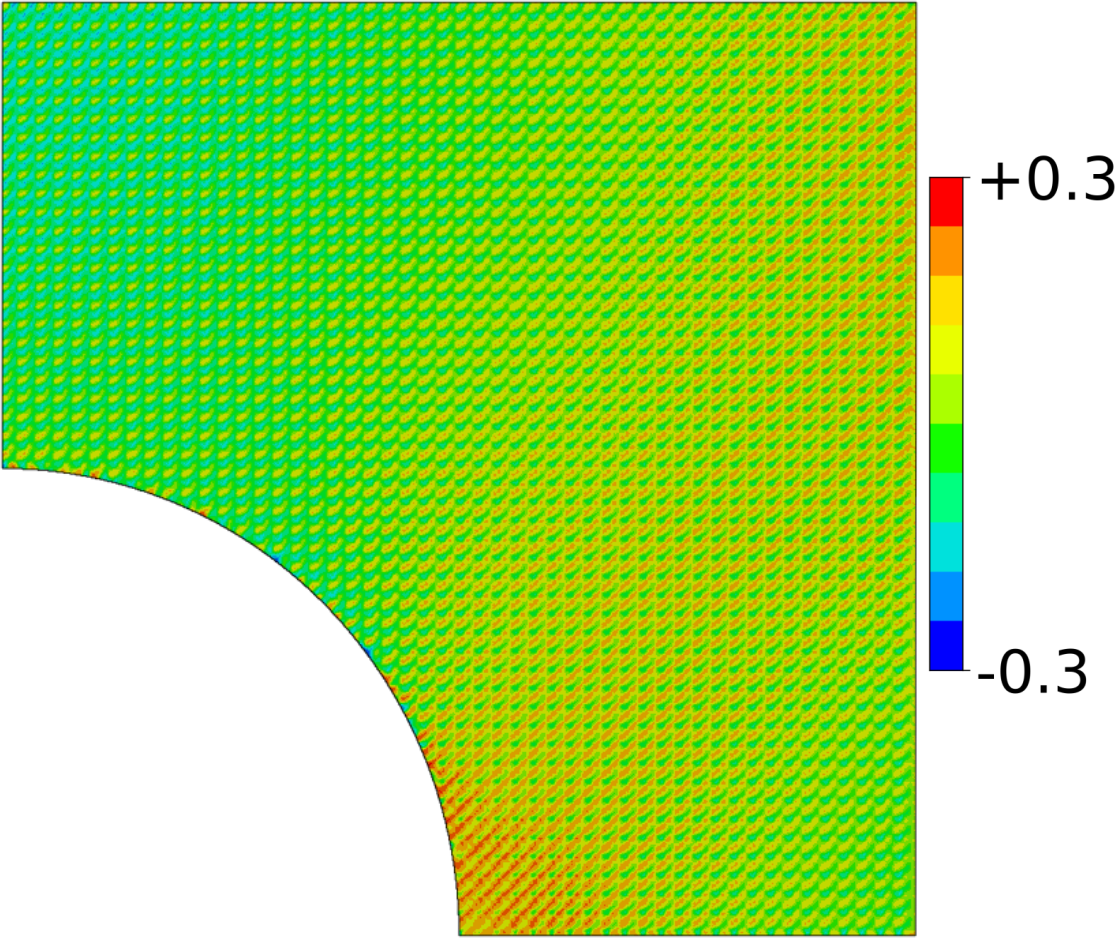}
            {\small $t/\tau=2.0$}
        \end{subfigure}
    
        \begin{subfigure}[b]{0.3\textwidth}
            \centering
            \includegraphics[width=\linewidth]{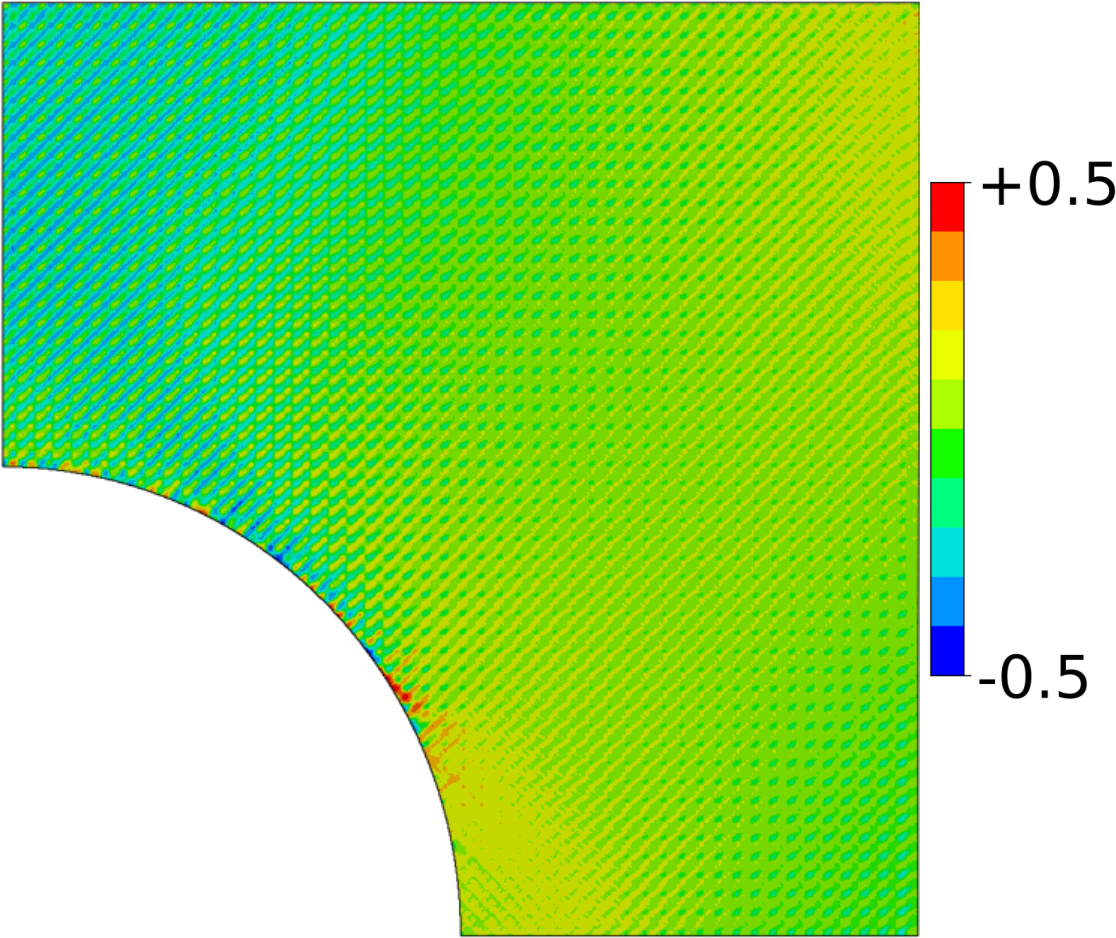}
            {\small $t/\tau=2.9$}
        \end{subfigure}
        \hfill
        \begin{subfigure}[b]{0.3\textwidth}
            \centering
            \includegraphics[width=\linewidth]{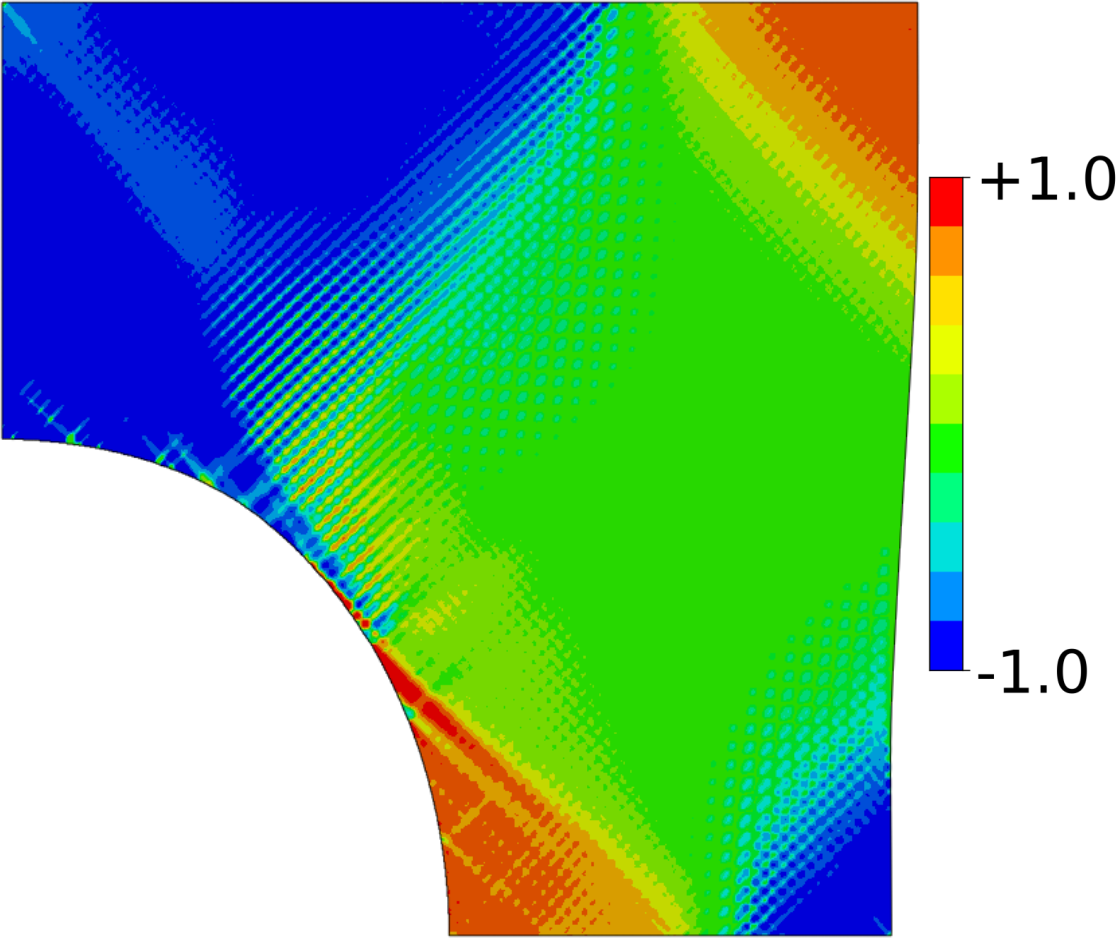}
            {\small $t/\tau=4.7$}
        \end{subfigure}
        \hfill
        \begin{subfigure}[b]{0.3\textwidth}
            \centering
            \includegraphics[width=\linewidth]{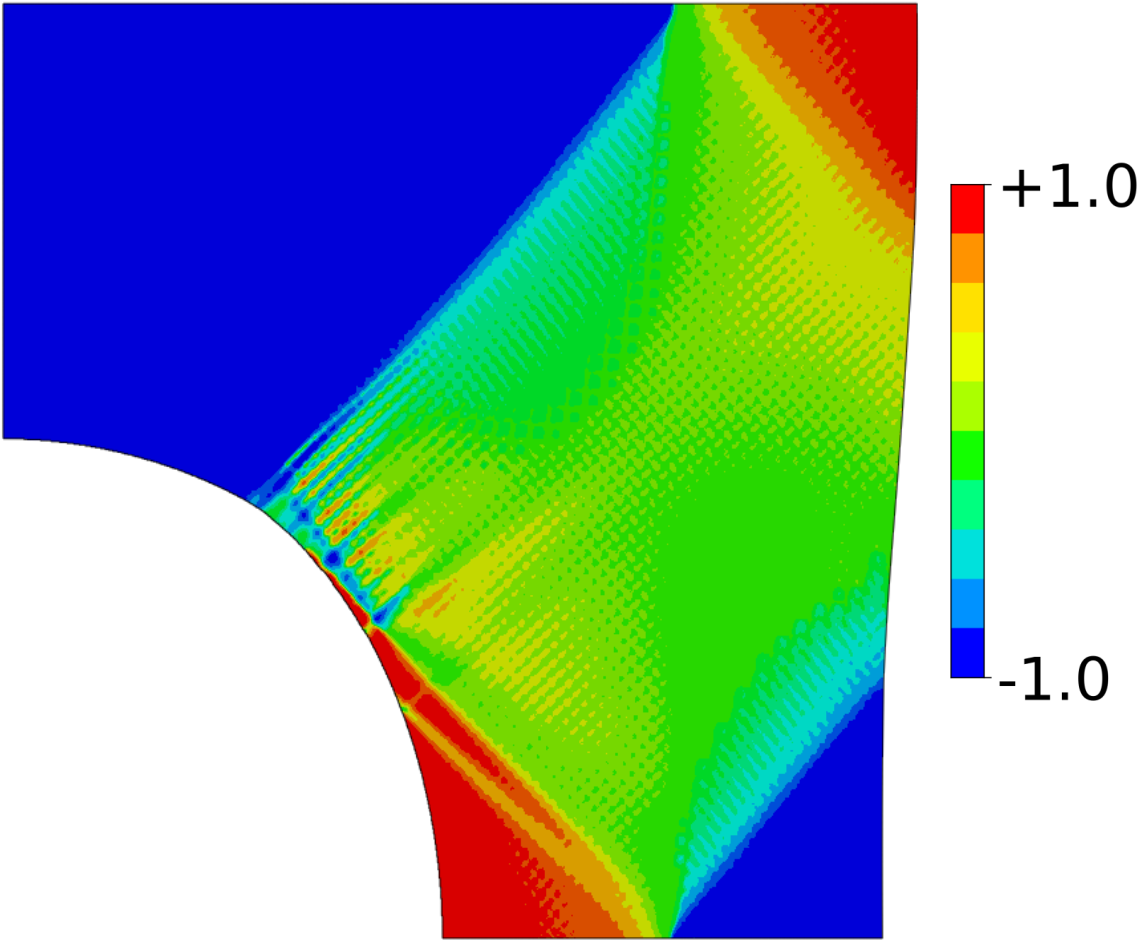}
            {\small $t/\tau=5.6$}
        \end{subfigure}
        \captionsetup{justification=centering}
        \caption{Direct numerical simulation}
         \label{fig:coupledDNS}
    \end{subfigure}

    \begin{subfigure}[b]{\textwidth}
        \centering
        \begin{subfigure}[b]{0.3\textwidth}
            \centering
            \includegraphics[width=\linewidth]{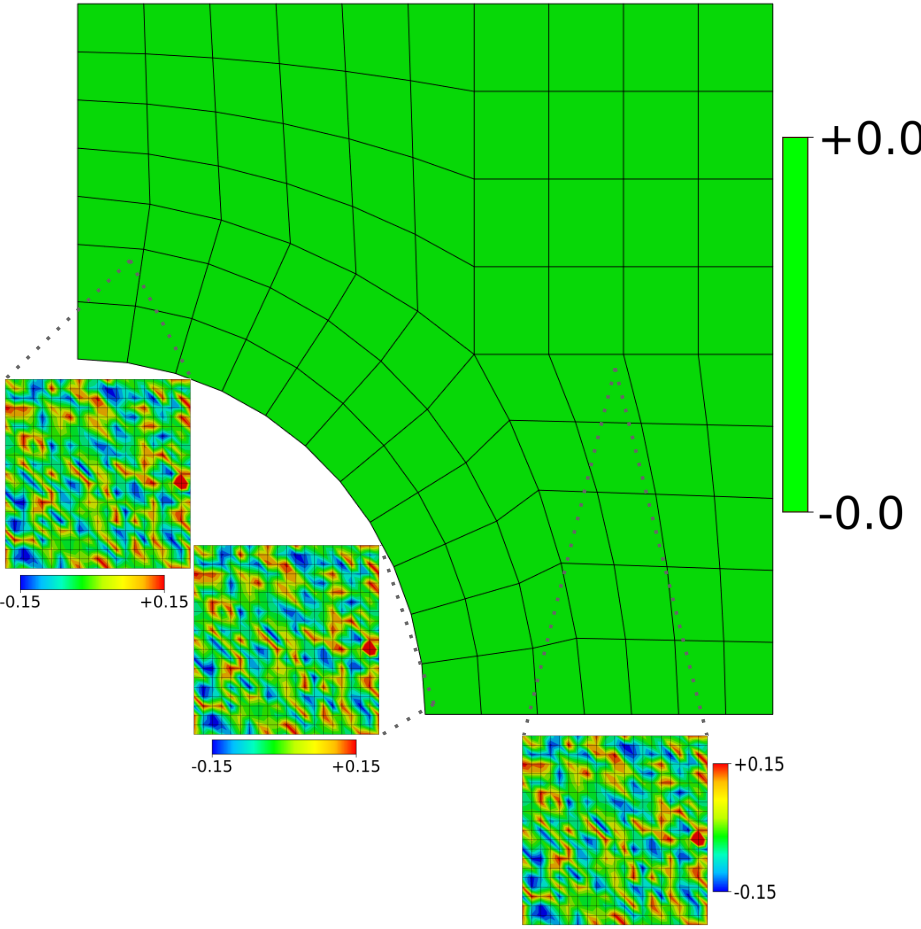}
            {\small $t=0$}
        \end{subfigure}
        \hfill
        \begin{subfigure}[b]{0.3\textwidth}
            \centering
            \includegraphics[width=\linewidth]{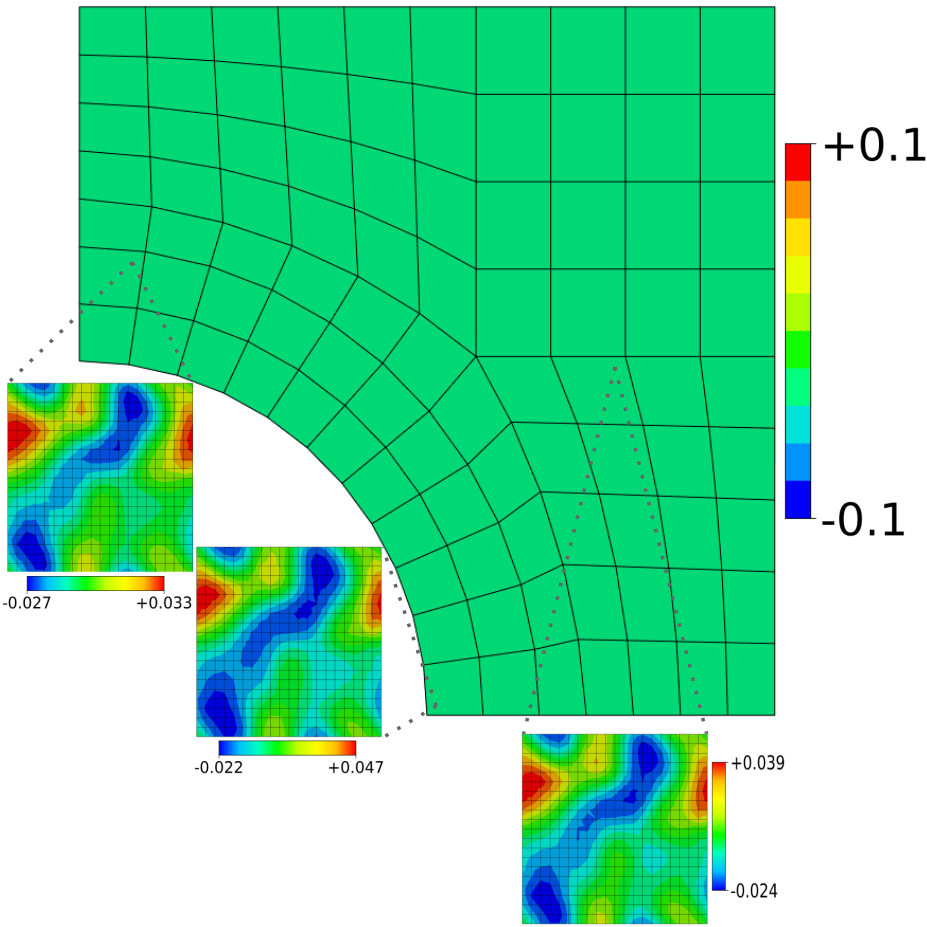}
            {\small $t/\tau=0.6$}
        \end{subfigure}
        \hfill
        \begin{subfigure}[b]{0.3\textwidth}
            \centering
            \includegraphics[width=\linewidth]{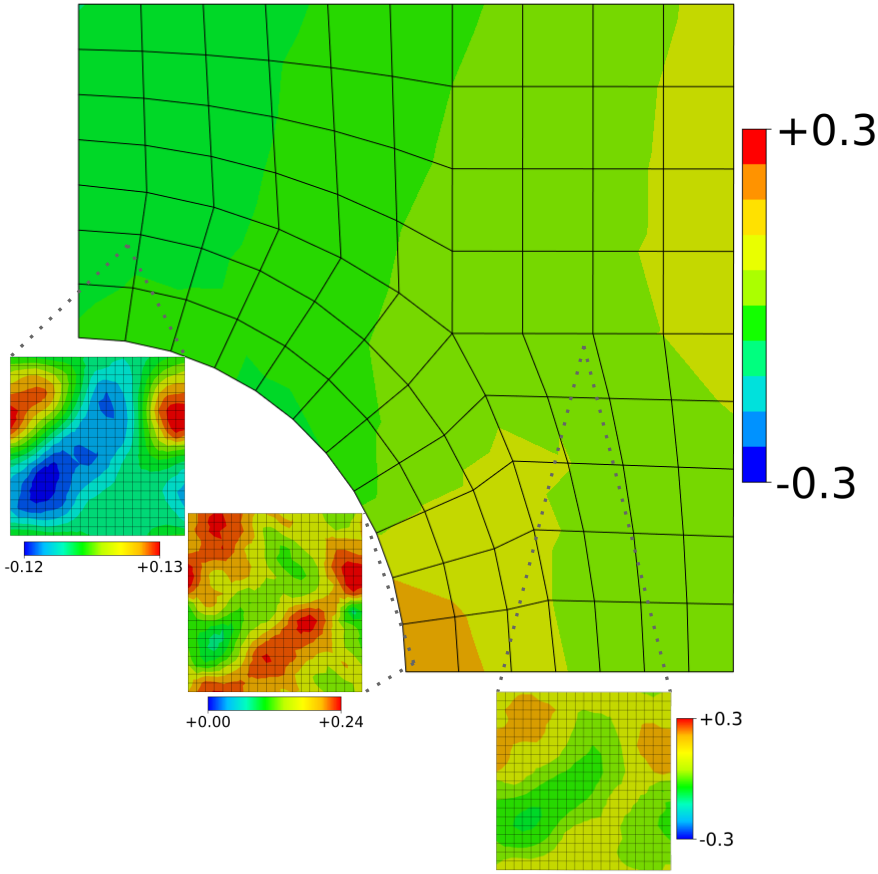}
            {\small $t/\tau=2.0$}
        \end{subfigure}
    
        \begin{subfigure}[b]{0.3\textwidth}
            \centering
            \includegraphics[width=\linewidth]{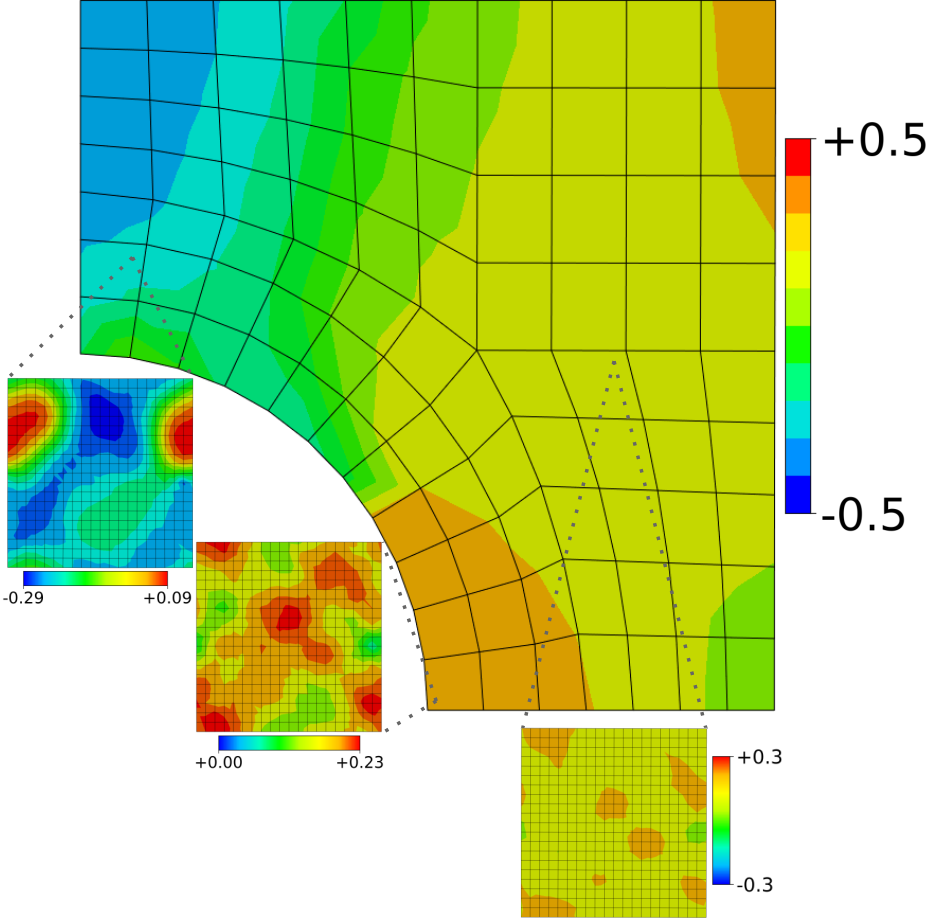}
            {\small $t/\tau=2.9$}
        \end{subfigure}
        \hfill
        \begin{subfigure}[b]{0.3\textwidth}
            \centering
            \includegraphics[width=\linewidth]{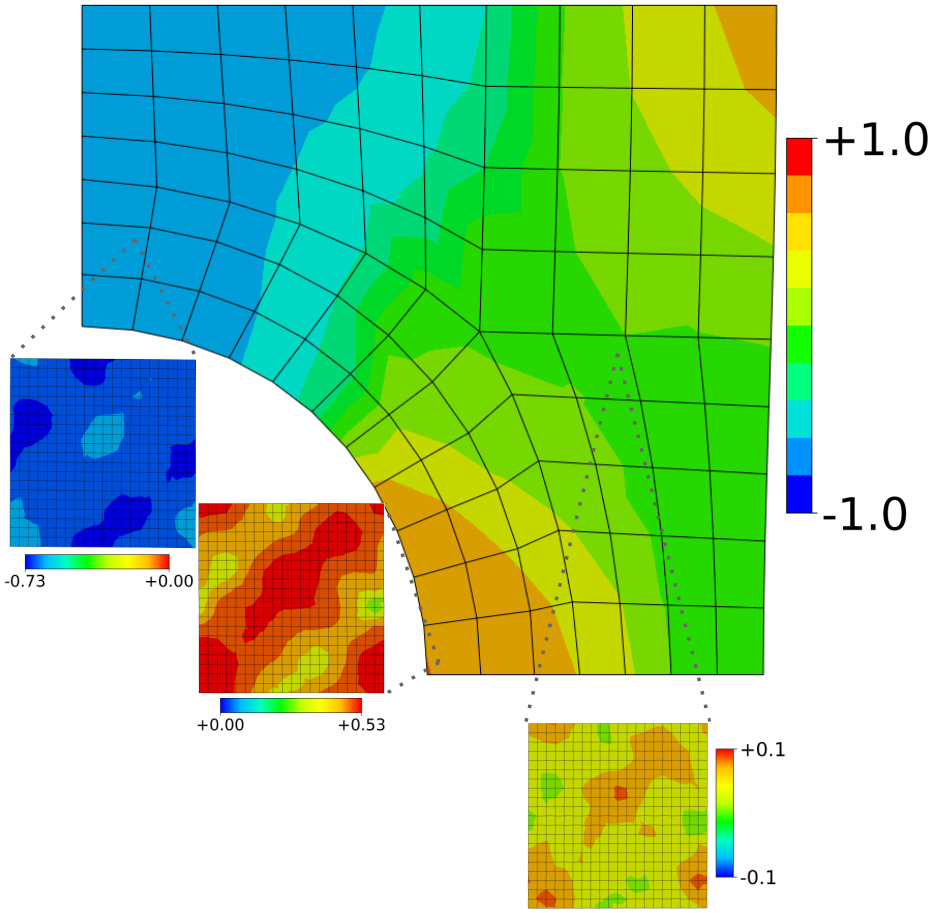}
            {\small $t/\tau=4.7$}
        \end{subfigure}
        \hfill
        \begin{subfigure}[b]{0.3\textwidth}
            \centering
            \includegraphics[width=\linewidth]{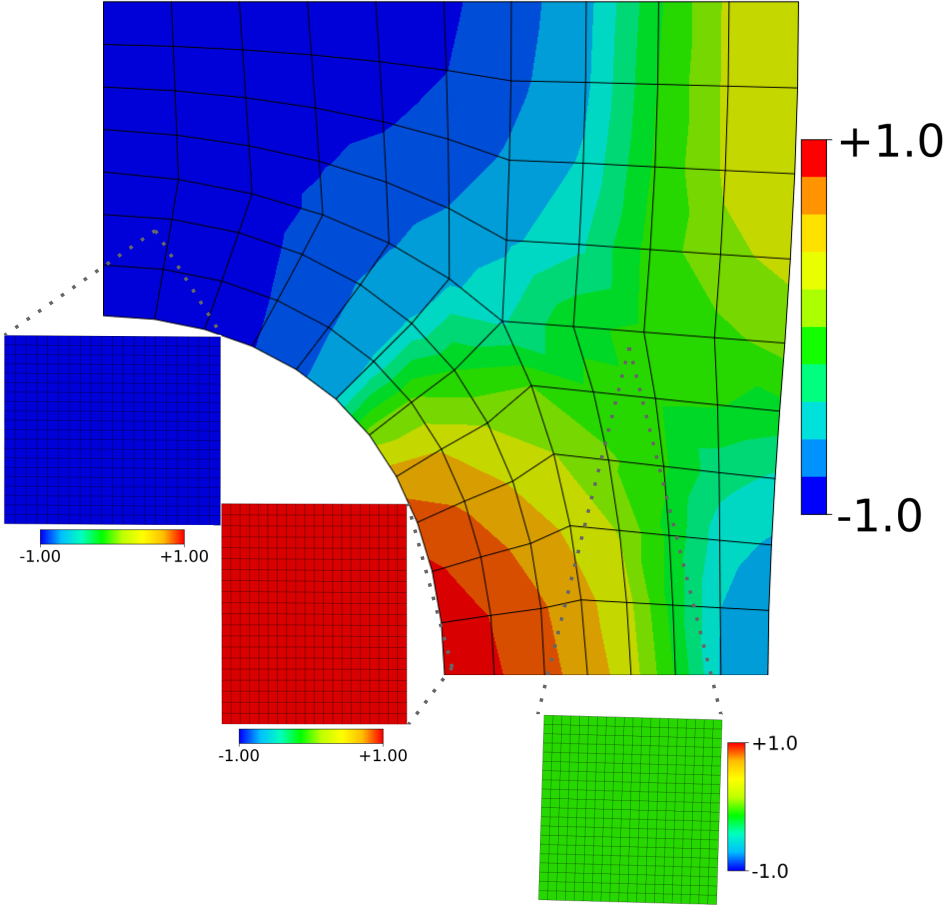}
            {\small $t/\tau=5.6$}
        \end{subfigure}
        \captionsetup{justification=centering}
        \caption{FE\textsuperscript{2} simulation }
        \label{fig:coupledFE2}
    \end{subfigure}
\caption{Comparison of the martensitic phase evolution in a plate with a hole under uniaxial tension, using a single-crystal RVE. Top row: Direct Numerical Simulation (DNS). Bottom row: FE\textsuperscript{2} simulation. The crystallographic axes are rotated by 45° relative to the loading direction. The color map represents the phase distribution, from martensitic variant $-1$ (blue) to austenite $0$ (green) to variant $+1$ (red).}
\label{fig:coupledplatewhole} 
\end{figure}

The results for the phase evolution from both the DNS and the FE\textsuperscript{2} simulation are shown in \figurename~\ref{fig:coupledplatewhole}. 
First, we examine the DNS results in \figurename~\ref{fig:coupledDNS}. The simulation begins with a consolidation of the random initial field into larger, more connected domains. As the mechanical loading increases, the stress concentration around the hole drives the transformation. The martensitic variants $\orderparam\!=\!-1$ and $\orderparam\!=\!1$ are favored and appear near the hole boundary in the regions where compressive and tensile hoop stresses, respectively, are predicted by Kirsch's analytical solution. Initially, a band of the parent austenite phase ($\orderparam\!\approx\!0$) remains along the diagonal of the model. As the loading progresses, this band finally transforms into the martensite variant $\orderparam\!=\!1$, with the transformation initiating from the top-right corner.

The FE\textsuperscript{2} results, presented in \figurename~\ref{fig:coupledFE2}, recover this complex behavior with reasonable accuracy, capturing both the spatial distribution of the phases and the temporal evolution of the transformation pattern. As expected and desired from a homogenization approach, the macroscopic field $\orderparammacro$ is inherently smoother than the microscopic field $\orderparam$ observed in the DNS. A final important note concerns the interpretation of the macroscopic fields, particularly in the last snapshot on the right-hand side of \figurename~\ref{fig:coupledFE2}. A value of $\orderparammacro\!\approx\!0$ could, in principle, represent either a region of homogeneous austenite or a finely twinned martensitic microstructure whose average resolves to zero. However, an inspection of the underlying microfields in the corresponding RVEs reveals that it is indeed untransformed austenite, confirming the predictive capability of the model.

A brief note on the computational costs further highlights the practical benefit of the proposed framework. For the plate with a hole example, the direct numerical simulation (DNS) required a solver time (evaluated as accumulated wall-clock time reported by Abaqus) of approximately \SI{20600}{\second}. In contrast, the corresponding FE\textsuperscript{2} simulation was completed in only \SI{4100}{\second}. Both simulations were run on a single compute node of our university's HPC cluster, equipped with 16 cores. This represents a computational speed-up of a factor of five, which is a significant achievement. It is particularly noteworthy given that this speed-up was obtained under conditions of low scale separation and without employing any further measures of improvement, such as model order reduction.

\section{Summary and Conclusions}

In this work, a comprehensive theory for the homogenization of phase-field models has been proposed. The framework establishes rigorous micro-macro relations for both the order parameter and its gradient. By formulating a condition of micro-homogeneity of the Hill-Mandel type in terms of Gurtin's microforces, compatibility is enforced in an average sense. The result is a well-posed boundary value problem (BVP) at the microscale, including the three classical types of boundary conditions known from computational micromechanics. The rigorous formulation of this BVP allows for the use of any suitable numerical technique for solving partial differential equations, such as the finite difference method (FDM), fast Fourier transforms (FFT), the finite volume method (FVM), or the finite element method (FEM).

For this study, the framework was exemplarily implemented within an FE\textsuperscript{2} scheme, employing periodic boundary conditions for the RVE. The capabilities of the proposed method were demonstrated through two distinct examples: a minimal Allen-Cahn model with a double-well potential, and a more complex, mechanically-coupled model for stress-driven martensitic phase transformation taken from the literature. Initial single RVE simulations were conducted to investigate the influence of the boundary condition type and to validate the fundamental predictions of the theory. Furthermore, since phase-field theory contains an intrinsic material length scale, first investigations into the effect of scale separation were performed. Finally, full FE\textsuperscript{2} simulations of notched structures demonstrated that the proposed framework operates reliably and allows for the prediction of the spatial and temporal evolution of the macroscopically \emph{smoothened} phase-field with reasonable accuracy, \emph{without} the need of a fully-resolved direct numerical simulation.

The choice of the RVE size in the present work has been rather pragmatic. In phase-field theory it essentially acts as an averaging window, which limits the gradients to be resolved at the macroscopic scale. Broader and more systematic investigations are required to establish robust guidelines for its selection. For many practical applications, such as polycrystalline materials, the grain size itself provides a relevant physical length scale on which the size of the RVE can be based.

In the investigated examples, which featured only a moderate separation of scales, the immediate computational gain of the FE\textsuperscript{2} simulations compared to fully-resolved direct numerical simulations (DNS) was limited. In fact, for the uncoupled phase-field example with the lowest scale separation, the FE\textsuperscript{2} simulation even took longer than the corresponding DNS, whereas for the mechanically-coupled example with higher scale separation a slight computational gain could be achieved. However, an immediate speed-up was neither to be expected nor within the scope of this work; its primary contribution is the establishment of a consistent homogenization framework.

This framework opens the door to the application of advanced numerical techniques to the microscale problem. For instance, while model order reduction \citep{Karatzas2021,LopezQuiroga2018,Yabansu2017} and machine learning-based surrogate models \citep{Wight2020,McClenny2023,Chen2025} have been widely applied to single-scale phase-field problems (see also the recent review by \citet{Lanzoni2025}), the proposed framework facilitates their use at the microscale in a multiscale setting. This has the potential to deliver speed-ups of several orders of magnitude and, ultimately, to enable phase-field simulations of complete engineering structures in future applications.

Finally, the proposed first-order phase-field homogenization framework offers several natural directions for future extensions. A straightforward generalization is the transition from a single scalar order parameter to multiple order parameters. Beyond that, more sophisticated enrichments, in the spirit of \citet{Forest2005Encyclopedia} for the mechanical case, could be pursued, i.e., higher-gradient generalizations or higher-order theories of micromorphic type; the latter may be realized, for example, by incorporating additional phase-morphology descriptors as recently  proposed in \cite{Oertzen2024}.


\section*{Acknowledgments}

The authors acknowledge computing time on the compute cluster of the Faculty of Mathematics and Computer Science of TU Bergakademie Freiberg, operated by the computing center (URZ) and funded by the Deutsche Forschungsgemeinschaft (DFG) under DFG grant number 397252409.

\section*{CRediT authorship contribution statement}

Geralf Hütter: Writing -- Original Draft, Methodology, Conceptualization. 
Fenil Lathiya:  Software, Investigation.
Vincent von Oertzen: Software, Writing –- Review \& Editing.
Bjoern Kiefer: Supervision, Writing –- Review \& Editing.

\section*{AI declaration statement }

During the preparation of this work, the authors used ScienceOS to support language editing and drafting, and to assist with identifying relevant literature. The authors reviewed and edited the content as needed, verified the results where appropriate, and take full responsibility for the content of the publication.

\printbibliography 

\end{document}